\documentstyle[11pt]{article}

\font\Bbbfont=msbm10
\def\Bbb#1{\hbox{\Bbbfont#1}}

\newcommand{\ra}{\rangle}
\newcommand{\la}{\langle}

\begin{document}

\hfill NBI-HE-98-30

\hfill DFNT-T 06-98

\hfill  MPS-RR-98-10

\begin{center}
{ \Large \bf Crumpled Triangulations
and
Critical Points\\
in
$4D$ simplicial quantum gravity}

\vspace{24pt}

{\large
{\sl J. Ambj\o rn}$\,^{a,}$\footnote{email ambjorn@nbi.dk. Supported
by a
MaPhySto-grant},
{\sl M. Carfora}$\,^{b,}$\footnote{email carfora@pv.infn.it;
carfora@sissa.it},
{\sl D. Gabrielli}$\,^{c,}$\footnote{email gabri@sissa.it}
and
{\sl A. Marzuoli}$\,^{b,}$\footnote{marzuoli@pv.infn.it}
}
\vspace{24pt}

$^a$~The Niels Bohr Institute, \\
Blegdamsvej 17, DK-2100 Copenhagen \O , Denmark,

\vspace{12pt}
$^b$~Dipartimento di Fisica Nucleare e Teorica, \\
Universit\`a degli Studi di Pavia, \\
via A. Bassi 6, I-27100 Pavia, Italy, \\
and\\
Istituto Nazionale di Fisica Nucleare, Sezione di Pavia, \\
via A. Bassi 6, I-27100 Pavia, Italy

\vspace{12pt}
$^c$~S.I.S.S.A.-I.S.A.S.,\\
Via Beirut 2-4, 34013 Trieste, Italy

\end{center}

\vspace{12pt}

\begin{center}
{\bf Abstract}
\end{center}

\vspace{12pt}
\noindent
We estimate analytically the critical coupling separating the weak
and the strong coupling regime in $4D$ simplicial quantum gravity to
be
located at $k_2^{crit}\simeq 1.3093$. By
carrying out a detailed geometrical analysis of the strong coupling
phase we argue that the distribution of dynamical triangulations with
singular vertices and singular edges, dominating in such a regime, is
characterized by
distinct sub-dominating peaks. The presence of such peaks generates
volume dependent pseudo-critical points:
$k_2^{crit}(N_4=32000)\simeq1.25795$,
$k_2^{crit}(N_4=48000)\simeq1.26752$,
$k_2^{crit}(N_4=64000)\simeq1.27466$, etc., which appear in good
agreement with available Monte Carlo data. Under a certain scaling
hypothesis we analytically characterize
the (canonical) average value, $c_1(N_4;k_2)=<N_0>/N_4$, and the
susceptibility, $c_2(N_4;k_2)=(<N_0^2>-<N_0>^2)/N_4$, associated with
the vertex distribution
of the $4$-D triangulations considered. Again, the resulting
analytical expressions are found in quite a good agreement with their
Monte Carlo counterparts.

\vfill

\newpage

\section{Introduction}

In this article we shall characterize analytically the critical point
separating the weak and the strong coupling regime in $4D$-simplicial
quantum gravity by locating it at $k_2^{crit}\simeq1.3093$. The
elementary techniques we develop here will allow us to get a rather
detailed understanding of the geometry and the physics of the strong
coupling phase of the theory. In particular we will show that the
dynamics in such a phase is influenced by the presence of peaks in the
distribution of singular triangulations. These latter are
combinatorial manifolds characterized by the presence of vertices
shared by a number of simplices diverging linearly with the volume of
the triangulation, and possibly connected by a sub-singular edge.
The peaks in question are parameterized by the fraction of total
volume which is allocated around such singular vertices. In order of
decreasing entropic relevance, the peaks are found, according to a
well-defined pattern, at
$k_2\simeq1.24465$, $k_2\simeq1.2744$, $k_2\simeq1.2938$,
$k_2\simeq1.30746$, $k_2\simeq1.31762$, $k_2\simeq1.32545$, etc.,
asymptotically fading towards the weak coupling regime. By exploiting
simple entropic arguments drawing from our recent work \cite{Carfora},
\cite{Houches} and  by making use of a certain scaling hypothesis, we
show how such collection of sub-dominating sets of singular
triangulations significantly affects the dynamics of the transition
between weak and strong coupling. We hope that our work offers the
possibility of making progress in understanding the nature of the
transition, one of the major issue which controls the validity of
dynamical triangulations as the basis of a regularization scheme for
gravity. Before embarking on this analysis we offer some general
motivation for
such a study.

\subsection{The Model}

Let $M$ be a closed n-dimensional, ($n\geq 2$), manifold of given
topology.
Let $Riem(M)$ and $Diff(M)$ respectively denote the space of
Riemannian  metrics $g$ on $M$, and the group of
diffeomorphisms on $M$.  In the
continuum formulation of Euclidean quantum
gravity one attempts to give meaning to a formal path integration over
$Diff(M)$
equivalence
classes of metrics in $Riem(M)$:
\vskip 0.5 cm
\begin{eqnarray}
Z(\Lambda, G,M)=\int_{Riem(M)/Diff(M)}{\cal
D}[g(M)]e^{-S_{g}[\Lambda,G,\Sigma]}
\label{uno}
\end{eqnarray}
\vskip 0.5 cm
\noindent where $S_{g}[\Lambda,G,\Sigma]$ is
the Einstein-Hilbert action associated with
the Riemannian manifold $(M,g)$, {\it viz.},
\vskip 0.5 cm
\begin{eqnarray}
S_g[\Lambda,G,\Sigma]=
\Lambda\int_M d^n\xi\sqrt{g}-\frac{1}{16\pi{G}}\int_{M}
d^n\xi\sqrt{g}\, R
\end{eqnarray}
\vskip 0.5 cm
\noindent and ${\cal D}[g(M)]$ is some {\it a priori} distribution on
$Riem(M)/Diff(M)$ describing the strong coupling statistics
($\Lambda\to0$, $G\to\infty$) of the set of Riemannian manifolds
$\{(M,g)\}$ considered. We avoid here discussing well-known specific
pathologies in dealing with (\ref{uno}) and about which the
reader can find abundant literature, and simply recall that in the
Dynamical Triangulations approach to quantum gravity one attempts to
give meaning to (\ref{uno}) by
replacing the continuum Riemannian
manifold $(M,g)$ with a Piecewise-Linear manifold (still denoted by
$M$) endowed with a triangulation $T_a\to{M}$ generated by gluing
a (large) number of equilateral $n$-simplices
$\sigma^n$. One approximates Riemannian structures
by means of such triangulated
manifolds by using a representative metric where each simplex
$\sigma^n$ is a Euclidean equilateral simplex with sides of length
$a$, (typically we set $a=1$). This metric is locally Euclidean
everywhere on the $PL$-manifold except near the $(n-2)$ sub-simplices
$\sigma^{n-2}$, (the {\it bones}), where
the sum of the dihedral angles, $\theta(\sigma^n)$, of the incident
$\sigma^n$'s is in excess (negative curvature) or in defect (positive
curvature) with respect to the $2\pi$ flatness constraint, the
corresponding deficit angle $r$ being defined by
$r=2\pi-\sum_{\sigma^n}\theta(\sigma^n)$. If $K^{n-2}$ denotes the
$(n-2)$-skeleton of $T\to{M^n}$, then $M^n\backslash{K^{n-2}}$ is a
flat Riemannian manifold, and any point in the interior of an $r$-
simplex $\sigma^r$ has a neighborhood homeomorphic to
$B^r\times{C}(link(\sigma^r))$, where $B^r$ denotes the ball in ${\Bbb
R}^n$ and ${C}(link(\sigma^r))$ is the cone over the link
$link(\sigma^r)$, (the product $link(\sigma^r)\times[0,1]$ with
$link(\sigma^r)\times\{1\}$ identified to a point). Note that for
dynamical triangulations the deficit angles are generated by the
string of integers, the {\it curvature assignments},
$\{q(k)\}_{k=0}^{N_{n-2}-1}$  providing the
numbers of top-dimensional simplices incident on the $N_{n-2}$
distinct bones, {\it viz.},
$r(i)=2\pi-q(i)\arccos(1/n)$.

\vskip 0.5 cm
By specializing to this setting the standard Regge calculus,
the formal path integration (\ref{uno}) is replaced on a
dynamically triangulated PL manifold $M$, (of fixed
topology), by the (grand-canonical) partition function
\cite{Regge},\cite{Houches},\cite{Frohlich}
\begin{eqnarray}
Z[k_{n-2},k_n]=\sum_{T\in {\cal T}(M)} \frac{1}{C_T}\;
e^{-k_nN_n+k_{n-2}N_{n-2}}
\label{grandpartition}
\end{eqnarray}
where $k_{n-2}$ and $k_n$ are two (running) couplings,
the former proportional to the inverse
gravitational coupling $1/G$, while the latter is a linear combination
of
$1/16\pi{G}$ and of the cosmological constant $\Lambda$.
The summation in (\ref{grandpartition}) is extended to the set
$\{{\cal T}(M)\}$ of all
distinct dynamical triangulations the PL-manifold $M$ can support, and
it is weighted by
the symmetry factor, $C_T$, of the triangulation: the
order of the automorphism group  of the graph associated with the
triangulation $T$. Since symmetric triangulations are the exception
rather then the rule, we shall assume $C_T=1$ in the estimates of
the partition functions below. Thus, in the following we will
omit the symmetry factor when writing the partition function.
\vskip 0.5 cm

One can introduce also the {\it canonical} partition
function  defined
by
\begin{eqnarray}
W(k_{n-2})_{eff}=\sum_{T\in {\cal T}(N_n)}
e^{k_{n-2}N_{n-2}},
\label{can}
\end{eqnarray}
where the summation is extended over all distinct dynamical
triangulations with given $N_n$, ({\it i.e.}, at fixed volume), of a
given PL-manifold $M$. Finally, we shall consider the
{\it micro-canonical} partition function
\begin{eqnarray}
W[N_{n-2},b(n,n-2)]=\sum_{T\in {\cal T}(N_n;N_{n-2})} 1,
\end{eqnarray}
where the summation is extended over all distinct dynamical
triangulations with given $N_n$ and $N_{n-2}$, {\it i.e.}, at fixed
volume and fixed {\it average incidence}
\begin{eqnarray}
b(n,n-2)=\frac{1}{2}n(n+1)(N_n/N_{n-2}),
\end{eqnarray}
of a given PL-manifold $M$.
The micro-canonical partition function is simply the number of
distinct dynamical triangulations with given volume ($\propto{N_n}$)
and fixed average curvature ($\propto{b(n,n-2)}$), of a given PL-
manifold $M$. In other words,  $W[N_{n-2},b(n,n-2)]$ is the {\em
entropy
function} for the given set of dynamical triangulations: it provides
the discretized counterpart of the a priori distribution ${\cal
D}[g(M)]$ describing the strong coupling statistics ($k_n\to0$,
$k_{n-2}\to0$) of the set of Riemannian manifolds $\{(M,g)\}$.
\vskip 0.5 cm

Recently there have been a number of significant advances in $3$- and
$4$-dimensional simplicial quantum gravity
that fit together in a coherent whole; roughly speaking,
these results are related to: (i) a deeper understanding of the
geometry of $n\geq3$-dimensional
dynamical triangulations; (ii) the study of simpler models mimicking
quite
accurately the critical structure of simplicial quantum gravity; (iii)
more refined computer simulations of the phase structure of the $4$-
dimensional theory. These results imply that
both in dimension $n=3$ and $n=4$, simplicial quantum gravity has two
geometrically
distinct phases parameterized by the value of the inverse
gravitational
coupling $k_{n-2}$. In the weak coupling phase (large values of $k_{n-
2}$) we have a dominance of PL-manifolds which collapse to branched
polymer structures with an Hausdorff dimension $d_H=2$ and an entropy
exponent (analogous to the string susceptibility of the $2$-
dimensional theory) $\gamma=1/2$. In this phase the theory has a well
defined continuum limit which is independent of any fine tuning of the
(inverse) gravitational constant $k_{n-2}$. We are not really
interested in this continuum limit, even if it exists. The situation
is not unusual from the point of view of lattice theories. For
instance in compact $U(1)$ gauge theories one hits the trivial Coulomb
phase for all $\beta>\beta_0$.
In the strong
coupling phase (small values of $k_{n-2}$), we have a dominance of
crumpled manifolds: the typical configuration sampled by the computer
simulations is
that of a triangulation with a few vertices on which most of the top-
dimensional simplices are incident, such presence of {\it singular
vertices}, typically connected by a sub-singular edge, seems to be a
signature
of the strong coupling phase \cite{Catterall}. Note
that the word {\it singular vertex} is used in DT theory with a
meaning quite different from the accepted meaning adopted in PL
geometry. What is actually meant is that
a metric ball around any such a vertex,
of radius equal to the given lattice spacing, has a volume that grows
proportionally to the volume of the whole PL-manifold.
This behavior indicates that the Hausdorff dimension of the typical
triangulation in the strong coupling phase is very large if not
infinite.
There is strong evidence that the transition between
weak and strong coupling, marked by a critical value $k_{n-
2}^{crit}$, is of a first order nature  in the $n=3$-dimensional case.
In dimension $n=4$, the original numerical simulations seemed to
indicate
a second order nature of the transition, a result that invited
positive speculations on the possibility that simplicial quantum
gravity could indeed provide a reliable regularization of Euclidean
quantum gravity. However, recent and more accurate analyses
\cite{Krzywicki} of the
Monte Carlo simulations seem
rather to point toward a first order nature of the transition. These
results are not definitive since the latent heat at the critical point
is very small as compared to the $3$D-case, so that the question
remains open whether any
further increase in the sophistication of the simulations will
definitively
establish, within the limit of the reached accuracy, the nature of the
transition. In any case, all recent numerical results \cite{singedge}
strongly indicates that the phase transition in $4$-dimensional
simplicial gravity is associated with the creation of singular
geometries.
These results
have put to the fore the
basic problems of theory, and in this sense important questions
abound: What is the geometrical nature of the crumpled phase? What is
the mechanism driving the transition from the polymer phase to the
crumpled phase? Is there any geometrical intuition behind the first or
higher order nature of the transition? Is it possible to take
dynamical control over the occurrence of singular vertices and edges
which
otherwise would entropically dominate?

Some of these questions can be systematically addressed by a detailed
but otherwise elementary discussion of the geometry of dynamical
triangulations along the lines of \cite{Carfora}, \cite{Houches}. This
geometrical
viewpoint has turned out to be useful and interesting in terms of
providing an analytical framework within which
discuss simplicial quantum gravity while at the same time maintaining
a strong contact with  computer simulations. In this paper we discuss
in detail how this approach can be further exploited to estimate
analytically the value of $k_2^{crit}$, and describing the
geometrical properties of the strong coupling phase of the $4$-
dimensional theory.

\section{Large volume asymptotics of the partition
functions}

For the convenience of the reader we recall in this section a few
basic results of \cite{Carfora} that we are going to use later on.

The discretized distribution $W[N_{n-2},b(n,n-2)]$ is one of the
objects of main interest in simplicial quantum gravity, and for
$n\geq3$ an exact evaluation of $W[N_{n-2},b(n,n-2)]$
is an open and very difficult problem. However,
since in (\ref{grandpartition}) and (\ref{can}) one is interested in
the large volume limit, what really matters, as far
as the criticality properties of (\ref{grandpartition}) are concerned,
is the asymptotic behavior of
$W[N_{n-2},b(n,n-2)]$ for large $N_n$.
This makes the analysis of $W[N_{n-2},b(n,n-2)]$ somewhat
technically  simpler, and according to
\cite{Carfora} one can actually estimate its
leading asymptotics with the relevant  sub-leading corrections.
If we consider an $n$-dimensional ($n\geq2$) PL-manifold $M$ of given
fundamental group $\pi_1(M)$), then
the distribution $W[N_{n-2},b(n,n-2)]$
of distinct dynamical triangulations, with given $N_{n-2}$ bones
and average curvature $b\equiv b(n,n-2)$, factorizes
according to
\begin{eqnarray}
W[N_{n-2},b(n,n-2)]=p_{N_{n-2}}^{curv} \la
Card\{T_a^{(i)}\}_{curv}\ra,
\label{rulesums}
\end{eqnarray}
where $p_{N_{n-2}}^{curv}$ is the number of possible
distinct curvature assignments $\{q(\alpha)\}_{\alpha=0}^{N_{n-2}}$
for triangulations
$\{T_a\}$ with $N_{n-2}$ bones and given average incidence $b(n,n-2)$,
{\it viz.},
\begin{eqnarray}
\{q(\alpha)\}_{\alpha=0}^{N_{n-2}}\not=
\{q(\beta)\}_{\beta=0}^{N_{n-2}}
\not=\{q(\gamma)\}_{\gamma=0}^{N_{n-2}}\not=\ldots,
\end{eqnarray}
while $<Card\{T_a^{(i)}\}_{curv}>$ is the average (with respect to the
distinct curvature assignments) of the number of distinct
triangulations sharing a common set of curvature assignments,
(for details, see section 5.2 pp.102 of \cite{Carfora}). This
factorization
allows a rather straightforward asymptotic analysis of $W[N_{n-
2},b(n,n-2)]$, and in the limit of large $N_n$ we get\cite{Carfora}
(theorem 6.4.2, pp. 131
with the expression $<Card\{T_a^{(i)}\}_{curv}>$ provided by theorem
5.2.1, pp. 106)
\begin{eqnarray}
W[N_{n-2},b(n,n-2)]&\simeq&
\frac{W_{\pi}}{\sqrt{2\pi}}\cdot
e^{(\alpha_nb(n,n-2)+\alpha_{n-2})N_{n-2}}
\nonumber\\
&&\cdot\sqrt{\frac{(b-\hat{q}+1)^{1-2n}}{(b-\hat{q})^3}}
\cdot
{\left [
\frac{(b-\hat{q}+1)^{b-\hat{q}+1}}{(b-\hat{q})^{b-\hat{q}}}
\right ] }^{N_{n-2}}
\label{asintotica}\\
&&e^{[-m(b(n,n-2))N^{1/n_H}_n]}
\left(\frac{b(n,n-2)}{n(n+1)}N_{n-2}\right)^{D/2}
{N_{n-2}}^{\tau(b)-\frac{2n+3}{2}}.
\nonumber
\label{notation}
\end{eqnarray}
\noindent The notation here is the following:\par
\noindent $W_{\pi}$ is a topology dependent parameter of no importance
for our
present purposes (see \cite{Carfora} for its explicit expression),
$\alpha_{n-2}$ and $\alpha_n$ are two constants depending on the
dimension $n$, (for instance, for $n=4$,
$\alpha_n=-\arccos(1/n)|_{n=4}$, $\alpha_{n-2}=0$); $\hat{q}$ is the
minimum incidence order over the bones (typically $\hat{q}=3$);
$D{\doteq}dim[Hom(\pi_1(M),G)]$ is the topological dimension of the
representation variety parameterizing the set of distinct dynamical
triangulations approximating locally homogeneous $G$-geometries,
($G\subset SO(n)$). Finally,
$m(b)\geq0$ and $\tau(b)\geq0$ are two parameters depending on
$b(n,n-2)$ which, together with $n_H>1$, characterize the sub-leading
asymptotics of
$W[N_{n-2},b(n,n-2)]$. In particular, note that
\begin{eqnarray}
e^{[-m(b(n,n-2))N^{1/n_H}_n]}
(\frac{b(n,n-2)}{n(n+1)}N_{n-2})^{D/2}
{N_{n-2}}^{\tau(b)}.
\label{asicurv}
\end{eqnarray}
\noindent is the asymptotics associated with
$\la Card\{T^{(i)}\}_{curv}\ra $.
The remaining part of (\ref{asintotica}) is the leading exponential
contribution coming from the large $N_n$ behavior of the distribution
$p_{N_{n-2}}^{curv}$ of the possible curvature assignments. This
latter term provides the correct behavior of the large volume limit of
dynamically triangulated manifolds, an asymptotics that
matches nicely with the existing Monte Carlo simulations, ({\it e.g.},
see \cite{Carfora}, section 7.1, pp.160).
\vskip 0.5 cm
While the exponential asymptotics is basically under control, it must
be stressed that some of
the most delicate aspects of the theory are actually contained in
$\la Card\{T^{(i)}\}_{curv} \ra$. Roughly speaking, the set of
triangulations $\{T^{(i)}\}_{curv}$ can be rather directly interpreted
as a finite dimensional approximation of the {\it moduli space} of
constant curvature metrics in smooth Riemannian geometry, (the
standard example being the parameterization of the moduli space of
surfaces of genus $\geq2$ with the set of inequivalent constant
curvature $(=-1)$ metrics admitted by such surfaces; due to rigidity
phenomena, such example is to be taken with care for $n\geq3$). With
this geometric interpretation in mind
we can consider
\begin{eqnarray}
\frac{\ln \la Card\{T^{(i)}\}_{curv} \ra}{\ln{N_n}}
{\simeq} -
\frac{m(b(n,n-2))N_n^{1/n_H}}{\ln{N_n}}+
\frac{D}{2}+\tau(n)~~~~~\mbox{for}~N_n\to \infty
\label{covdim}
\end{eqnarray}
as the (formal) covering dimension of such a moduli space. Clearly,
whenever $m(b)>0$ (and
$n_H$ finite or $O(\ln{N_n})$), such covering dimension is singular.
This signals the fact that
in the corresponding range of the parameter $b$, dynamical
triangulations fail to approximate in the large volume limit any
smooth Riemannian manifold.

\subsection{The critical incidence}
\label{critica}
 The parameters $m(b)$, $\tau(b)$,
and
$n_H$, characterizing the covering dimension and the large volume
asymptotics (\ref{asicurv})
are not yet explicitly provided by the analytical results of
\cite{Carfora}. By exploiting geometrical arguments, one can only
prove\cite{Carfora} (theorem 5.2.1, pp. 106) an existence result to
the effect
that  if
$n\geq3$, there is a {\it critical value}
$b_0(n)$, of the average incidence  $b(n,n-2)$,
to which we
can associate  a critical value
$k_{n-2}^{crit}$ of the inverse gravitational coupling,
such that
\begin{eqnarray}
m(b)=0,
\label{muno}
\end{eqnarray}
for $b(n,n-2)\leq b_0(n)$; whereas
\begin{eqnarray}
m(b)>0,
\label{mdue}
\end{eqnarray}
for $b_0(n)< b(n,n-2)$. In other words, for $b<b_0(n)$ the
sub-leading asymptotics in (\ref{asintotica})
is at most polynomial, whereas for $b>b_0(n)$ this asymptotics
becomes sub-exponential as $N_n$ goes to infinity,
(note that in the $2$-dimensional case (\ref{asintotica}) has always a
sub-leading
polynomial asymptotics).

This change in the sub-leading asymptotics qualitatively accounts for
the jump from the strong to the weak coupling phase observed in the
real system during Monte Carlo simulations. However, the lack of an
explicit
expression for $m(b)$ hampers a deeper analysis of the nature of this
transition. In particular, one is interested in
the way the
parameter $m(b(n,n-2))$ approaches $0$ as $b(n,n-2)\to{b_0(n)}$, since
adequate knowledge in this direction would provide the order of the
phase transition.  It is clear that a first necessary step in
order to discuss the properties of $m(b(n,n-2))$ is to provide a
constructive geometrical characterization of the critical average
incidence $b_0(n)$, and not just an existence result.

As far as the other parameter $\tau(n)$ is concerned, the situation is
on more firm ground. $\tau(n)$ characterizes the sub-leading
polynomial
asymptotics in the weak coupling phase, and recently\cite{Gabrielli},
an analysis of the geometry of dynamical triangulations
in this phase has provided
convincing analytical evidence that $\tau(n)-(2n+3)/2+3=1/2$.
As expected, this corresponds to a dominance, in the weak coupling
phase, of branched polymers structures.

\subsection{Canonical Averages and the Curvature Susceptibility}
\label{canone}
If we consider the weighted distribution $W[N_{n-2},b(n,n-
2)]\exp[k_{n-2}N_{n-2}]$, characterizing the canonical partition
function (\ref{can}), then it is straightforward to check that, as
$N_n\to\infty$, this distribution is strongly peaked around
triangulations with an average incidence $b^*(n,n-2;k_{n-2})$ given by
(see \cite{Carfora},
eqn. (6.39), pp. 134)
\begin{eqnarray}
b^*(n,n-2;k_{n-2})=3(\frac{A(k_{n-2})}{A(k_{n-2})-1}),
\label{solution}
\end{eqnarray}
\noindent where for notational convenience we have set
\begin{eqnarray}
A(k_{n-2})&\doteq&
{ \left [\frac{27}{2}e^{k_{n-2}}+1+
\sqrt{(\frac{27}{2}e^{k_{n-2}}+1)^2-1} \right ]}^{1/3}+
\nonumber\\
&&{ \left [\frac{27}{2}e^{k_{n-2}}+1-
\sqrt{(\frac{27}{2}e^{k_{n-2}}+1)^2-1} \right ]}^{1/3}-1.
\end{eqnarray}
This remark allows to compute, via a uniform Laplace estimation, the
large volume asymptotics of the canonical partition function $W(k_{n-
2})_{eff}=\sum_{T\in {\cal T}(N_n)}
e^{k_{n-2}N_{n-2}}$. A discussion of this asymptotics at the various
orders is rather delicate and the reader can find the details in
\cite{Carfora}, (chapter 6, theorem 6.6.1). For our purposes it is
sufficient to
consider the leading order expression which can be readily obtained,
starting from the micro-canonical partition function, by means of a
standard saddle point evaluation, {\it viz.}
\begin{eqnarray}
W(N_n, k_{n-2})_{eff}&=&
c_n\left( \frac{(A(k_{n-2})+2)}{3A(k_{n-2})} \right)^{-n}
N_n^{\tau(n)+D/2-n-
1}e^{[-m(b^*(n,n-2;k_{n-2}))N^{1/n_H}_n]}
\cdot\nonumber\\
&&\cdot e^{\left[[\frac{1}{2}n(n+1)\ln\frac{A(k_{n-2})+2}{3}]N_n
\right]}\Big(1+O(N^{-3/2}_n)\Big),
\label{regular}
\end{eqnarray}
where $c_n$ is a scaling factor not depending from $k_2$.

\vskip 0.5 cm
As the inverse gravitational coupling $k_{n-2}$ varies, the
average curvature correspondingly changes according to
(\ref{solution}). It follows that there is a well defined critical
value,
$k_{n-2}^{crit}$, solution of the equation
\begin{eqnarray}
b_0(n)=3(\frac{A(k_{n-2})}{A(k_{n-2})-1}),
\label{criteqn}
\end{eqnarray}
for which $b^*(n,n-2;k_{n-2})=b_0(n)$, where $b_0(n)$ is the critical
average incidence, (see (\ref{muno}) and (\ref{mdue})).
This $k_{n-2}^{crit}$ describes
the transition between the strong coupling phase ($k_{n-2}<k_{n-
2}^{crit}$) and
the weak coupling phase ($k_{n-2}>k_{n-2}^{crit}$) associated with the
two distinct sub-leading asymptotics regimes of
(\ref{regular}).
\vskip 0.5 cm
 The explicit geometric characterization of
$b_0(n)$ and  the evaluation of the corresponding critical value
$k_{2}^{crit}$ in the $4$-dimensional case are among the most
important issues that we discuss in this paper. In order to compare
the geometrical results we obtain with the data coming from recent
Monte Carlo simulation it will be useful to have handy the expressions
of the free energy $\ln{W_{eff}(N_4,k_2)}$, of
the (canonical) average of the number of bones
$<N_2>=\partial\ln{W_{eff}(N_4,k_2)}/\partial{k_2}$, and of the
associated {\it curvature-curvature} correlator $[\la N_2^2\ra-
\la N_2\ra^2]=\partial^2\ln{W_{eff}(N_4,k_2)}/\partial{k_2}^2$.
\vskip 0.5 cm
The large volume asymptotics of the (canonical) free energy is readily
obtained from (\ref{regular}); by setting $n=4$ and discarding the
inessential constant terms we get (in the saddle-point approximation
used above)
\begin{eqnarray}
\ln{W(N, k_{2})_{eff}}=
10N_4\ln\frac{A(k_{2})+2}{3}-m(k_2)N^{1/n_H}.
\label{frenergy}
\end{eqnarray}
\vskip 0.5 cm
The canonical average $<N_2>$ follows from differentiating
(\ref{frenergy}) with respect to the inverse gravitational coupling
$k_2$,
\begin{eqnarray}
\la N_2\ra =10\frac{A_1(k_2)}{A(k_2)+2}N_4-
N_4^{1/n_H}\frac{\partial{m(k_2)}}{\partial{k_2}},
\label{firstderiv}
\end{eqnarray}
where we have set
\begin{eqnarray}
 A_1(k_2) &\doteq& \frac{\partial{A(k_2)}}{\partial{k_2}}\nonumber\\
&= &\frac{9}{2}e^{k_2}
\left[1+\frac{\frac{27}{2}e^{k_{2}}+1}{\sqrt{(\frac{27}{2}e^{k_{2}}+1)
^2-1}}\right]\cdot
{ \left [\frac{27}{2}e^{k_{2}}+1+
\sqrt{(\frac{27}{2}e^{k_{2}}+1)^2-1} \right ]}^{-2/3}+
\nonumber\\
&& +\frac{9}{2}e^{k_2}
\left[1-
\frac{\frac{27}{2}e^{k_{2}}+1}{\sqrt{(\frac{27}{2}e^{k_{2}}+1)^2-
1}}\right]\cdot
{ \left [\frac{27}{2}e^{k_{2}}+1-
\sqrt{(\frac{27}{2}e^{k_{2}}+1)^2-1} \right ]}^{-2/3}.
\end{eqnarray}
\vskip 0.5 cm
Note that for a $4$-dimensional PL manifold $M$ of Euler
characteristic $\chi(M)$, we have $N_0=\frac{N_2}{2}-N_4+\chi$. Thus,
from $\la N_2\ra$ we immediately get the expression for
the first normalized cumulant of the distribution of the number of
vertices of the triangulation, {\it viz.}
\begin{equation}
c_1(N_4;k_2)\doteq\frac{\la N_0\ra }{N_4}=
\frac{5A_1(k_2)}{A(k_2)+2}-1 -N_4^{1/n_H-
1}\frac{\partial{m(k_2)}}{\partial{k_2}}
\label{cumulant1}
\end{equation}
\vskip 0.5 cm
This is a typical quantity monitored in Monte Carlo simulation, and
later on we will discuss how
(\ref{cumulant1}) actually compares with respect existing numerical
data.

Finally, the curvature susceptibility
\begin{eqnarray}
\frac{\la N_2^2\ra - \la N_2\ra^2}{N_4}
=\frac{1}{N_4}\; \frac{\partial^2\, \ln{W_{eff}(N_4,k_2)}}{\partial{k_
2}^2}
\end{eqnarray}
is explicitly computed as
\begin{eqnarray}
4c_2(N_4;k_2)&=& \frac{\la N_2^2\ra-\la N_2\ra^2}{N_4}\nonumber\\
&=& 10\frac{(A(k_2)+2)A_2(k_2)-A_1(k_2)^2}{(A(k_2)+2)^2}
-N_4^{1/n_H-1}\frac{\partial^2{m(k_2)}}{\partial{k_2}^2},
\label{cumulant2}
\end{eqnarray}
\noindent where $c_2(N_4;k_2)\doteq\frac{\la N_0^2\ra-\la
N_0\ra^2}{N_4}$
is the second normalized cumulant of the distribution of the number of
vertices of the triangulation, and where we have set
\begin{eqnarray}
&& A_2(k_2)\doteq\frac{\partial^2{A(k_2)}}{\partial{k_2}^2}=
A_1(k_2)+\nonumber\\
&& + \frac{\frac{243}{4}e^{2k_{2}}}
{[(\frac{27}{2}e^{k_{2}}+1)^2-1]^{3/2}}{
\left[\frac{27}{2}e^{k_{2}}+1-
\sqrt{(\frac{27}{2}e^{k_{2}}+1)^2-1} \right]}^{-2/3}-\nonumber\\
&& - \frac{\frac{243}{4}e^{2k_{2}}}
{[(\frac{27}{2}e^{k_{2}}+1)^2-1]^{3/2}}{\left[\frac{27}{2}e^{k_{2}}+1+
\sqrt{(\frac{27}{2}e^{k_{2}}+1)^2-1}  \right]}^{-2/3} -\nonumber\\
&& -\frac{81}{2}e^{2k_2}{ \left[\frac{27}{2}e^{k_{2}}+1+
\sqrt{(\frac{27}{2}e^{k_{2}}+1)^2-1} \right]}^{-
5/3}\left[1+\frac{\frac{27}{2}e^{k_{2}}+1}{\sqrt{(\frac{27}{2}e^{k_{2}
}+1)^2-1}}\right]^2-\nonumber\\
&& -\frac{81}{2}e^{2k_2}{\left[\frac{27}{2}e^{k_{2}}+1-
\sqrt{(\frac{27}{2}e^{k_{2}}+1)^2-1} \right]}^{-5/3}\left[1-
\frac{\frac{27}{2}e^{k_{2}}+1}{\sqrt{(\frac{27}{2}e^{k_{2}}+1)^2-
1}}\right]^2
\label{der2}
\end{eqnarray}

\subsection{A scaling hypothesis}
Clearly the above expressions for $c_1(N_4;k_2)$ and $c_2(N_4;k_2)$
are useless if we do not specify how
$m(k_2)$ depends on the inverse gravitational coupling $k_2$.
Since according to theorem 5.2.1 of \cite{Carfora},
$m(k_2)\to0$ as $b(k_2)$ approaches a critical incidence
$b_0(4)$, (henceforth denoted by $b_0$), the simplest {\it hypothesis}
we can make is that, for
$(b(k_2)-b_0)\to0^+$, $m(k_2)$ scales to zero according to a power law
given by
\begin{equation}
m(k_2)=\frac{1}{\nu}(\frac{1}{b(k_2)}-\frac{1}{b_0})^{\nu},
\label{assumption}
\end{equation}
where $0<\nu<1$ is a critical exponent to be determined; (the factor
$1/\nu$ is inserted for later convenience).

The expression (\ref{regular}) for the canonical partition function
$W[N_{n-2}, b(n,n-2)]$ is a large volume ($N_4$) asymptotics evaluated
at fixed volume. It contains a non- trivial subleading asymptotics
governed by $m(k_2)$. The net effect of this subleading term is
clearly visible in the expressions (\ref{cumulant1}) and
(\ref{cumulant2}) of the two cumulants, and shows that in order to
capture the behavior of $c_1(N_4;k_2)$ and $c_2(N_4;k_2)$ as
$k_2\to{k_2}^{crit}$, (\ref{assumption}) is not sufficient. It must be
combined with a finite scaling hypothesis telling us how $m(k_2)$
scales with the volume $N_4$, as $b(k_2)\to{b_0}$. From the
asymptotics (\ref{regular}), and
the expression (\ref{cumulant1}) for the first cumulant, it easily
follows that the simplest, if not the most natural, hypothesis we can
make is to assume that $m(k_2)$ scales asymptotically with the volume
according to
\begin{eqnarray}
\lim_{ \matrix{{\scriptstyle N_4\to\infty}, \cr
{\scriptstyle (k_2-{k_2^{crit}})\to0^-}}}
|\frac{1}{b(k_2)}-\frac{1}{b_0}|^{\nu-1}\cdot{N_4}^{\frac{1}{n_H}-
1}=1,
\label{scaling}
\end{eqnarray}
where according to theorem 5.2.1 of \cite{Carfora} $n_H>1$.
This implies that, when $b(k_2)\to{b}_0$,
$m(k_2)$ scales as
\begin{eqnarray}
m(k_2)\simeq N_4^{-\frac{\nu(n_H-1)}{n_H(1-\nu)}}.
\end{eqnarray}
Together with $m(k_2)N_4^{1/n_H}\to0$ as $b(k_2)\to{b_0}$,
(\ref{scaling}) yields  $(1/n_H)<\nu<1$, a finite size scaling
relation connecting
the critical exponents $\nu$ and $n_H$.

Introducing this ansatz in (\ref{cumulant1}) and (\ref{cumulant2}) we
explicitly obtain
\begin{equation}\label{Rumulant1}
c_1(N_4;k_2)\doteq\frac{\la N_0 \ra}{N_4}
=\frac{5A_1(k_2)}{A(k_2)+2}-\frac{1}{3}\frac{A_1(k_2)}{A^2(k_2)}-1,
\end{equation}
and
\begin{eqnarray}
c_2(N_4;k_2)&=&\frac{\la N_0^2\ra-\la
N_0\ra^2}{N_4}\label{Rumulant2}\\
&=& \frac{5}{2}\frac{(A(k_2)+2)A_2(k_2)-A_1(k_2)^2}{(A(k_2)+2)^2}-
\frac{1}{12}\frac{A_2(k_2)A(k_2)-2A_1(k_2)}{A^3(k_2)}+\nonumber\\
&& +\frac{|\nu-1|}{36}\;\frac{A_1^2(k_2)}{A^4(k_2)}\;
{\left\vert \frac{A(k_2)-1}{3A(k_2)}-\frac{1}{b_0}  \right\vert}^{-1}.
\nonumber
\end{eqnarray}

Note that in this latter expression the only undetermined parameters
are the critical exponent $\nu$ and the critical incidence $b_0$. In
the following paragraphs we provide an explicit geometric
characterization of such $b_0$. The only remaining unknown quantity is
then $\nu$. Anticipating the conclusion of the paper, it
turns out that it is possible the choose a value
of  $\nu$ ($\approx 0.94$) which leads to
quite a good agreement between (\ref{Rumulant1})--(\ref{Rumulant2})
and the available Monte Carlo data for the
distributions of these two cumulants.

\section{The Geometry of the strong coupling phase}

A PL manifold endowed with a dynamical triangulation
is a particular example of an Alexandrov space, {\it i.e.} finite
dimensional, inner metric space with a lower curvature bound in
distance-comparison sense, (a brief introduction with the relevant
references can be found in \cite{Carfora}, section 3.2). The natural
topology
specifying in which sense dynamical triangulations approximate
Riemannian manifolds is associated with an Hausdorff-like distance
introduced by Gromov \cite{Gromov}, and which is a direct
generalization of the classical Hausdorff distance between compact
subsets of a metric space. The role of this topology stems from the
fact there are many geometric constructions in dynamical
triangulations theory that are close in Gromov-Hausdorff topology, but
not in smooth Riemannian geometry.
 In \cite{Carfora}  we proved that every Riemannian manifold (of
bounded geometry) can be uniformly approximated in this topology by
dynamical triangulations, (\cite{Carfora}, section 3.3, Th.3.3.1); the
converse result, namely if every dynamical triangulation approximates,
as the number of simplices goes to $\infty$, a $n$-dimensional
Riemannian manifold, is deeply tied to understanding the structure of
the thermo-dynamical behavior of the large volume limit of the set of
possible dynamical triangulations. It is interesting to note that the
geometry of the set of all possible dynamical triangulations of a
manifold of given topology is a subject of which little is actually
known, even in dimension two. Recently, in a remarkable paper
\cite{Thurston}, W. Thurston has shed some light in the two-
dimensional case by showing that the space of triangulations (of
positive curvature) has the rich geometric structure of a complex
hyperbolic manifold. We do not need to reach, in this work, such a
level of sophistication and
 in order to determine the critical average incidence, $b_0$, we
discuss mostly the kinematical properties of the
space of all possible dynamical triangulations admitted by an $n$-
dimensional PL manifold $M$ of given topology.
\vskip 0.5 cm
Let $(M,T_a)$ be a dynamically triangulated manifold, then
the $f$-
vector of the triangulation is the string of integers
$(N_0(T_a),N_1(T_a),\ldots,N_n(T_a))$,
where $N_i(T_a)\in {\Bbb N}$ is the number of $i$-dimensional sub-
simplices
$\sigma^i$ of $T_a$. This vector is constrained by the Dehn-
Sommerville relations
\begin{eqnarray}
\sum_{i=0}^n(-1)^iN_i(T)=\chi(T),
\label{eleven}
\end{eqnarray}
\begin{eqnarray}
\sum_{i=2k-1}^n(-1)^i\frac{(i+1)!}{(i-2k+2)!(2k-1)!}N_i(T)=0,
\label{twelve}
\end{eqnarray}
 if $n$ is even, and  $1\leq k\leq n/2$. Whereas if
$n$ is odd
\begin{eqnarray}
\sum_{i=2k}^n(-1)^i\frac{(i+1)!}{(i-2k+1)!2k!}N_i(T)=0,
\label{thirteen}
\end{eqnarray}
 with $1\leq k\leq (n-1)/2$, and where $\chi(T)$ is the Euler-
Poincar\'e characteristic of $T$.
 It is easily verified that the relations (\ref{eleven}),
(\ref{twelve}), (\ref{thirteen})
leave $\frac{1}{2}n-1$, ($n$
even) or $\frac{1}{2}(n-1)$ unknown quantities  among the $n$ ratios
$N_1/N_0,\ldots,N_n/N_0$, \cite{Dav4}. Thus, in dimension
$n=2,3,4$, the datum of $N_n$,  and
of the number of bones $N_{n-2}$, fixes, through the Dehn-Sommerville
relations
all the remaining $N_i(T)$. These extremely simple and perhaps even
naive-sounding remarks turn out to be quite powerful in providing
information on the global metrical
properties of the underlying PL-manifold. Not only, as is obvious, on
the volume
($\propto{N_n(T_a)}$),  and on the  average curvature
($\propto\frac{1}{2}n(n+1){N_n(T_a)/N_{n-2}(T_a)}$), but, corroborated
by a few more elementary
facts, also on the genesis of singular vertices and edges.
 \vskip 0.5 cm
 An
elementary but geometrically significant result of this type is
provided by the
range of variation of the possible average incidence $b(n,n-2)$. One
gets (see\cite{Carfora} (lemma 2.1.1))
\newtheorem{lemma}{Lemma}
\begin{lemma}
Let $T_a\to M^n$ a triangulation of a closed $n$-dimensional $PL$-
manifold $M$, with $2\leq{n}\leq4$, then for $N_n(T_a)\to\infty$, we
get\par
\noindent (i) For $n=2$:
\begin{eqnarray}
b(2,0)=6;
\end{eqnarray}
\noindent (ii) for $n=3$:
\begin{eqnarray}
\frac{9}{2}\leq b(3,1)\leq6;
\label{bitre}
\end{eqnarray}
\noindent (iii) for $n=4$:
\begin{eqnarray}
 4\leq b(4,2)\leq 5.
\label{biquattro}
\end{eqnarray}
\label{walkup}
\end{lemma}

The $2$-dimensional case as well as the upper bounds for
$n=3$ and $n=4$ are well-known trivial consequences of the Dehn-
Sommerville
relations. The lower bounds $b(3,1)\geq9/2$ and $b(4,2)\geq4$ are
instead related to a rather sophisticated set of results
proved by
Walkup\cite{Walkup} in the sixties concerning the proof of some
conjectures  for $3$-
and $4$-dimensional PL manifolds; (apparently, these results went
unnoticed by researchers in simplicial quantum gravity). Walkup's
theorems
have
important implications for understanding the geometry both of the
strong and of the weak coupling phase of simplicial gravity.
In
dimension $n=3$, we have\cite{Walkup}
\newtheorem{theorem}{Theorem}
\begin{theorem}
There exists a triangulation $T\to{\Bbb S}^3$ of the $3$-sphere ${\Bbb
S}^3$ with $N_0$ vertices and $N_1$ edges if and only if
$N_0\geq5$ and
\begin{eqnarray}
4N_0-10\geq N_1 \geq \frac{N_0(N_0-1)}{2}.
\end{eqnarray}
Moreover $T$ is a triangulation of ${\Bbb S}^3$ satisfying
$N_1=4N_0-10$ if and only if $T$ is a stacked sphere,
whereas $T$ is a triangulation of ${\Bbb S}^3$ satisfying
$N_1 =\frac{N_0(N_0-1)}{2}$ if and only if $T$ is a
$2$-neighborly
triangulation, namely if every pair of vertices is connected by an
edge.
\end{theorem}

A stacked sphere $({\Bbb S}^n,T)$ is a
triangulation $T\to{\Bbb S}^n$ of a sphere which
can be constructed from the boundary $\partial\sigma^{n+1}\simeq{\Bbb
S}^n$ of a simplex $\sigma^{n+1}$ by successive adding
of pyramids over some facets. More explicitly,
the boundary complex of any abstract $(n+1)$-simplex $\sigma^{n+1}$ is
by definition
a stacked sphere, and if $T$ is a stacked sphere and $\sigma^n$ is any
$n$-simplex of $T$, then $\hat{T}$ is a stacked sphere if $\hat{T}$ is
any complex obtained by $T$ by removing $\sigma^n$ and adding the join
of the boundary $\partial\sigma^n$ with a new vertex distinct from the
vertices of $T$. Note also that a triangulated $PL$-manifold is called
$k$-neighborly if
\begin{eqnarray}
N_{k-1}(T)=\frac{N_0!}{k!(N_0-k)!}
\end{eqnarray}

We are referring explicitly to $3$- and $4$-spheres ${\Bbb S}^n$,
because the majority of Monte Carlo simulations have been carried out
in these cases (for a recent discussion of more general topologies,
see\cite{Bialas}). However, it must be stressed that the above
definitions, as well as Walkup's theorems, can be naturally
extended (with suitable modifications\cite{Walkup})
to any $n$-dimensional PL manifolds $M$. Note in particular
that every triangulable $3$-manifold $M$ can be triangulated so that
the
closed star of some edge contains all the vertices and every pair of
vertices is connected by an edge.

In dimension $n=4$ we have a somewhat weaker characterization of the
possible set of triangulations:
\begin{theorem}
If $T\to{M}$ is a triangulation of a closed connected $4$-manifold,
then
\begin{eqnarray}
N_1(T)\geq 5N_0(T)-\frac{15}{2}\chi(T),
\label{fourstacked}
\end{eqnarray}
and equality holds if and only if $(M,T)$ is a stacked sphere.
\end{theorem}
 Note that actually one has a stronger statement in the sense that
equality in (\ref{fourstacked}) holds if and only is all vertex links
in the $4$-manifold $M$ are stacked $3$-spheres.

Contrary to what happens for $3$-manifolds, $2$-neighborly
triangulations ({\it i.e.}, triangulations where every pair of
vertices is connected by an edge), are not {\it generic}
for $4$-dimensional PL manifolds, and as matter of fact, the above
theorem
immediately implies\cite{Kuhnel} that for any such $(M,T)$
\begin{eqnarray}
N_0(T)(N_0(T)-11)\geq -15\chi(M),
\label{neigh}
\end{eqnarray}
where the equality implies that $(M,T)$ is $2$-neighborly. Thus, the
equality is not possible for large and arbitrary values of
$N_0(T)$, but, (depending on topology)\cite{Kuhnel}, only in the cases
$N_0(T)=0$, $N_0=5$, $N_0=6$, or $N_0=11\,mod\,15$.

Even if $2$-neighborly triangulations are not generic, one can easily
construct voluminous ({\it i.e.}, with $N_4(T)$ arbitrarily large)
triangulations of the $4$-sphere where all vertices {\it but two} are
connected by an edge. In order to realize such triangulations,
consider a
$2$-neighborly triangulation $T(3)$ of the $3$-sphere ${\Bbb S}^3$
with $f$-vector
$[N_0(T(3)),N_1(T(3)),N_2(T(3)),N_3(T(3))]$. If we take
the {\it Cone}, $C({\Bbb S}^3)$, on such $({\Bbb S}^3,T(3))$,
{\it viz.}, the product ${\Bbb S}^3\times[0,a]$ with
${\Bbb S}^3\times\{a\}$ identified to a point, then we get a
triangulation of a $4$-dimensional ball $B^4$ with $f$-vector
given by
\begin{eqnarray}
f(B^4)&= &(N_0(T(3))+1, N_1(T(3))+N_0(T(3)),\nonumber\\
&& N_2(T(3))+
N_1(T(3)), N_3(T(3))+N_2(T(3)), N_3(T(3))).
\end{eqnarray}
By gluing two copies of such a cone $C({\Bbb S}^3)$ along their
isometric
boundary $\partial{C}({\Bbb S}^3)\simeq{\Bbb S}^3$, we get a
triangulation of the $4$-sphere ${\Bbb S}^4$ with $f$-vector
\begin{eqnarray}
f({\Bbb S}^4) &= &
(N_0(T(3))+2, N_1(T(3))+2N_0(T(3)),\nonumber\\
&& N_2(T(3))+
2N_1(T(3)), N_3(T(3))+2N_2(T(3)), 2N_3(T(3))).
\end{eqnarray}
It is trivially checked that corresponding to such a triangulation we
get
\begin{eqnarray}
N_1({\Bbb S}^4)=\frac{N_0({\Bbb S}^4)(N_0({\Bbb S}^4)-1)}{2}-1,
\end{eqnarray}
where the $-1$ accounts for the missing edge between the two cone-
vertices in $C({\Bbb S}^3)\cup_{S^3}{C}({\Bbb S}^3)$.
\vskip 0.5 cm
When applied to simplicial quantum gravity, the existence of such $2$-
neighborly (or almost $2$-neighborly) triangulations implies that
there are dynamical triangulations of the $n$-sphere ${\Bbb S}^n$,
$n=3$, $n=4$, where all vertices are singular. Corresponding to such
configurations we have that $b(n,n-
2)|_{n=3}=6$, and $b(n,n-2)|_{n=4}=5$. Thus, not surprisingly,  for
such
triangulations the kinematical upper bound for
the average incidence $b(n,n-2)$ is attained. However, it is
important to stress that such extremely singular configurations {\it
do not saturate} the set of possible configurations for which
$b_{max}(n,n-2)$ is reached. From the Dehn-Sommerville relations one
immediately gets
\begin{eqnarray}
b(n,n-2)|_{n=3}=6\cdot\frac{N_3}{N_3+N_0},
\end{eqnarray}
and
\begin{eqnarray}
b(n,n-2)|_{n=4}=10\cdot\frac{N_4}{2N_4+2N_0-2\chi(T)},
\end{eqnarray}
which, together with the obvious relation $N_1\leq{N_0}(N_0-1)/2$,
implies that in order to attain the upper
kinematical bounds $b_{max}(n,n-2)|_{n=3}=6$ and
$b_{max}(n,n-2)|_{n=4}=5$ it is sufficient that
\begin{eqnarray}
N_0(T)=O[N_n(T)^{\alpha}],
\end{eqnarray}
with $1/2\leq\alpha<1$. Note that $2$-neighborly or almost $2$-
neighborly
triangulations correspond to $\alpha=1/2$.
\subsection{Singular Stacked Spheres}
\label{sss}
It should be stressed that the presence of singular vertices can occur
also for $b(n,n-2)=b_{min}(n,n-2)$, {\it i.e.}, for stacked spheres.
In other words, singular vertices are {\it not kinematically}
forbidden by the geometry of the triangulations. Their suppression or
enhancement in the different phases of simplicial quantum gravity is
rather related to the relative abundance, with respect to the
totality of possible triangulations, of the number of distinct
triangulations with singular vertices as $b(n,n-2)$ varies.
In other words, it is an entropic phenomenon as clearly suggested by
S. Catterall, G. Thorleifsson, J. Kogut, and R. Renken
\cite{Catterall}.
For definiteness, we
 can describe a concrete construction of a singular stacked sphere. It
amounts to gluing a $4$-dimensional ball $B^4$ bounded by a stacked
$3$-sphere ${\Bbb S}^3$ with a cone over such an ${\Bbb S}^3$.

\vskip 0.5 cm
Consider a $3$-dimensional stacked sphere ${\Bbb S}^3$. According to
one of Walkup's theorems, such an ${\Bbb S}^3 $ is the boundary of a
$4$-dimensional ball $B^4$ with a tree-like structure and
corresponds to a triangulation with $f$-vector
\begin{equation}
f(B^4) =  (N_0(S^3), N_1(S^3), N_2(S^3), N_3(S^3)+N_3(\hat{B}^4),
N_4(B^4)),
\end{equation}
where $N_i(S^3)$, $i=0,1,2,3$ is the $f$-vector of the boundary
stacked sphere and $N_3(\hat{B}^4)$ is the number of $\sigma^3$ in the
interior, $\hat{B}^4$, of $B^4$. Note that if we take the cone,
$C(S^3)$, over the boundary stacked sphere, we get another
triangulation of the $4$-dimensional ball, $B^4_{sing}$,
whose boundary is again isometric to the given ${\Bbb S}^3$, but whose
interior contains a (unique) singular vertex. The $f$-vector of such a
triangulation is
\begin{equation}
f(B^4_{sing})\! =\! (1\!+\! N_0(S^3), N_1(S^3)\!+\!N_0(S^3),
N_2(S^3)\!+\!N_1(S^3),
N_3(S^3)\!+\!N_2(S^3), N_3(S^3)).
\end{equation}
Gluing these two triangulated balls $B^4$ to $B^4_{sing}$ through
their common boundary ${\Bbb
S}^3$, we get a triangulation of the $4$-sphere
${\Bbb S}^4\simeq B^4\cup_{S^3}C(\partial{B^4})$, with $f$-vector
\begin{eqnarray}
N_0&= & N_0(S^3)+1\nonumber\\
N_1&= & N_1(S^3)+ N_0(S^3)\nonumber\\
N_2&= & N_2(S^3)+ N_1(S^3)\nonumber\\
N_3&= & N_3(S^3)+ N_3(\hat{B}^4)+N_2(S^3)\nonumber\\
N_4&= & N_4(B^4)+ N_3(S^3).
\end{eqnarray}
Since ${\Bbb S}^3$ is a stacked sphere, we have $4N_3(S^3)=3N_1(S^3)$
which, together with $N_2(S^3)=2N_3(S^3)$, implies
\begin{eqnarray}
N_2=\frac{10}{3}N_3(S^3).
\end{eqnarray}
From $2N_3=5N_4$ and the Euler relation for the $4$-dimensional ball
$B^4$, (with $\chi(B^4)=1$), we immediately get
$N_4(B^4)=\frac{1}{3}N_3(S^3)-\frac{2}{3}$, which implies
\begin{eqnarray}
N_4=\frac{4}{3}N_3(S^3)-\frac{2}{3}.
\end{eqnarray}
Thus, for $N_3(S^3)\to\infty$ we get a voluminous triangulation
($N_4\to\infty$) of ${\Bbb S}^4$ with average incidence
\begin{eqnarray}
b(n,n-2)|_{n=4}=10\cdot\frac{N_4=\frac{4}{3}N_3(S^3)-
\frac{2}{3}}{N_2=\frac{10}{3}N_3(S^3)}\to_{N_3\to\infty} 4,
\end{eqnarray}
which shows that ${\Bbb S}^4\simeq B^4\cup_{S^3}C(\partial{B^4})$ is a
stacked sphere with a singular vertex (the apex of the cone
$C(\partial{B^4})$).
An even simpler construction suffices to prove an analogous result in
the
$3$-dimensional case.
\vskip 0.5 cm
As mentioned in the introductory remarks, stacked spheres are relevant
in providing the geometrical rationale for the prevalence of branched
polymer structures in the weak coupling phase of simplicial quantum
gravity. As a matter of fact \cite{Gabrielli},it is their tree-like
structure that
accounts for the {\it kinematical} possibility of
polymerization. However, the existence of stacked spheres with
singular vertices, shows that the dynamical onset of polymerization is
not just a consequence of the geometry of triangulated manifolds as
$b(n,n-2)\to{b}_{min}(n,n-2)$.
On the kinematical side we may have, in the configuration space,
extremal cases such as the
$2$-neighborly triangulations
occurring for $b(n,n-2)\to{b}_{max}(n,n-2)$ or the singular stacked
spheres for
$b(n,n-2)\to{b}_{min}(n,n-2)$. Monte Carlo
simulations do confirm that such configurations are not generic. Near
$b_{max}(n,n-
2)$  we generically sample singular triangulations with just a few
singular vertices\cite{Catterall}. Similarly, as
$b(n,n-2)\to{b}_{min}(n,n-2)$ the dominant configurations sampled
correspond to stacked spheres without singular vertices. The mechanism
for understanding the dynamical prevalence of such configurations over
the other configurations which are kinematically possible is simply
related to
the fact that, with respect to the counting measure, distinct
dynamical triangulations are not equally probable as a function of the
average incidence $b(n,n-2)$. In order to discuss this point we need
to exploit a few elementary facts related to the geometry of the
ergodic moves used in simplicial quantum gravity.

\subsection{Ergodic moves and the onset of criticality}
\label{ergodic}
The $(k,l)$ moves\cite{Varsted} in $3$ and $4$ dimension are a well
known set of elementary surgery operations (related to the Pachner
moves\cite{Pachner}) which allow to construct all triangulations of a
PL- manifolds starting from a given triangulation. Roughly speaking,
the generic $(k,l)$ move consists in cutting out a sub-complex made up
of $k$-dimensional
simplices $\sigma^k$ and replacing it with a complex of $l$-
dimensional simplices $\sigma^l$ with the same boundary. Note that
$k+l=n+2$. We are interested in discussing how a finite set of such
moves generate the $f$-vector of voluminous triangulations of the $n$-
sphere ${\Bbb S}^n$, ($n=3,4$), starting from the standard $f$-vector
of the simplex $\partial\sigma^{n+1}\simeq{\Bbb S}^n$. For $n=3$, the
relevant moves are the $(1,4)$ move (barycentric subdivision), the
$(2,3)$ move (triangle to link exchange) and their inverses.
For $n=4$, since the {\it flip} move $(3,3)$ does not alter the
distribution of the number $N_i(T)$ of simplices,
the $f$-vector of the sphere is generated by the moves
$(1,5)$ (barycentric subdivision)  and $(2,4)$ (two-four exchange) and
their inverses.
Following\cite{Gabrielli}, (with a slight change in
notation), we denote by $P_{k,l}(n)$ the number of moves of type
$(k,l)$ in dimension $n$, and introduce the balance variables ($n=3$):
$x_1\doteq(P_{1,4}-P_{4,1})$, $x_2\doteq(P_{2,3}-P_{3,2})$;
and ($n=4$): $y_1\doteq(P_{1,5}-P_{5,1})$, $y_2\doteq(P_{2,4}-
P_{4,2})$. In terms of such quantities we can easily characterize the
string of integers $\{N_i\}$, $i=0,\ldots,n$, which are {\it possible}
$f$-vectors of triangulated ${\Bbb S}^n$.

For $n=3$, we get for $f({\Bbb S}^3) =
(N_0(S^3),N_1(S^3),N_2(S^3),N_3(S^3))$:
\begin{eqnarray}
N_0(S^3)&=& 5+x_1,\nonumber\\
N_1(S^3)&=&10+4x_1+x_2,\nonumber\\
N_2(S^3)&=&10+6x_1+2x_2,\nonumber\\
N_3(S^3)&=&5+3x_1+x_2,
\label{tredim}
\end{eqnarray}
whereas, for $n=4$ we have for
$f({\Bbb S}^4) = (N_0(S^4),N_1(S^4),N_2(S^4),N_3(S^4),N_4(S^4))$:
\begin{eqnarray}
N_0(S^4)&=&6+y_1,\nonumber\\
N_1(S^4)&=&15+5y_1+y_2,\nonumber\\
N_2(S^4)&=&20+10y_1+4y_2,\nonumber\\
N_3(S^4)&=&15+10y_1+5y_2,\nonumber\\
N_4(S^4)&=& 6+4y_1+2y_2.
\label{fourdim}
\end{eqnarray}
Note that not all $f({\Bbb S}^n)$ obtained in this way are actual $f$-
vectors of triangulated ${\Bbb S}^n$. This is a consequence of the
fact that the above relations between the $\{N_i\}$ and the variables
$P_{k,l}(n)$ are equivalent to the Dehn-Sommerville constraints. And
these latter are known to be necessary but not sufficient conditions
in characterizing the possible $f$-vectors of a triangulated manifold;
(sufficient conditions have been conjectured by R.
Stanley\cite{Stanley}-see\cite{Carfora} for a brief discussion of
this point).

Walkup's theorems imply the following kinematical bounds on the
variables $x_i$, $y_i$, ($i=1,2$):
\begin{eqnarray}
x_1 &\geq & 0,\nonumber\\
y_1 &\geq & 0,
\end{eqnarray}
(both from the obvious condition $N_0(S^n)\geq n+2$);
\begin{eqnarray}
x_2 &\geq & 0,\nonumber\\
y_2 &\geq & 0,
\end{eqnarray}
(the former from $N_1(S^3)\geq4N_0(S^3)-10$; the latter from
$N_1(S^4)\geq5N_0(S^4)-\frac{15}{2}\chi(S^4)$, with $\chi(S^4)=2$);
\begin{eqnarray}
x_1^2+x_1-2x_2 &\geq & 0,\nonumber\\
y_1^2+y_1-2y_2 &\geq & 0,
\end{eqnarray}
(both from $N_1(S^n)\leq{N_0}(S^n)(N_0(S^n)-1)/2$).
Finally, one can express the average incidence $b(n,n-2)$ as a
function of $x_i$ and $y_i$, so as to obtain
\begin{eqnarray}
b(n,n-2)|_{n=3}=6\cdot\frac{5+3x_1+x_2}{10+4x_1+x_2},
\end{eqnarray}
and
\begin{eqnarray}
b(n,n-2)|_{n=4}=10\cdot\frac{6+4y_1+2y_2}{20+10y_1+4y_2}.
\end{eqnarray}
\vskip 0.5 cm
It is also interesting to discuss in terms of the variables $x_i$ and
$y_i$, the average incidence of the top-dimensional simplices
$\sigma^n$ on the vertices $\sigma^0$ of the triangulations
considered. A straightforward computation provides
\begin{eqnarray}
Q(n)\doteq\frac{1}{N_0}\sum_{\{\sigma^0\}}q(\sigma^0)=
(n+1)\frac{N_n}{N_0},
\end{eqnarray}
yielding
\begin{eqnarray}
Q(n)|_{n=3}=4\cdot\frac{5+3x_1+x_2}{5+x_1},
\end{eqnarray}
and
\begin{eqnarray}
Q(n)|_{n=4}=5\cdot\frac{6+4y_1+2y_2}{6+y_1}.
\end{eqnarray}
As expected, $Q(n)$ is not bounded above: when the move $(2,3)$ (for
$n=3$), or $(2,4)$ (for $n=4$) dominates, {\it i.e.}, near the
$b_{max}(n,n-2)$ kinematical boundary, $Q(n)\to\infty$ as
$N_n\to\infty$. One may wonder if this unboundedness is related to the
unboundedness of the Einstein-Hilbert action, the answer is most
likely no. It is certainly reasonable to put restrictions on $Q(n)$ in
the search of a continuum limit of the theory, and this may change the
phase structure of the theory. But if it does, it is just an
illustration of the fact that this particular part of the phase
diagram  has no relevance for a genuine continuum limit. There should
be a reasonable universality. This is nicely illustrated in $2$-D
dynamical triangulation theory where any restriction (except the
strict flatness constraint
$q(i)=6$) leads to $2$D-gravity.

\vskip 0.5 cm

The above elementary remarks are a trivial restatement of the well
known
fact  that the moves $(1,4)$ and
$(1,5)$ (the barycentric subdivision) drive the system into the
elongated phase, whereas the moves $(2,3)$ and $(2,4)$ drive to the
crumpled phase. The  crumpling transition occur as soon as singular
vertices are
statistically enhanced by the presence of enough $(2,4)$ moves
with respect to $(1,5)$,
(for $n=3$ this enhancement is generated by the dominance of $(2,3)$
moves with respect to $(1,4)$ moves).

\subsection{The genesis of singular vertices: ${\Bbb S}^4_{sv}$}
\label{singvertex}
In order to characterize the onset of crumpling we describe the $f$-
vector of the generic triangulation of ${\Bbb S}^n$
in a way that clearly shows the mechanism of formation of singular
vertices. Such a description is obtained by gluing a triangulated
ball $B^n$ to the cone over its boundary $\partial{B}^n\simeq{\Bbb
S}^{n-1}$.
Thus, by referring to the $4$-
dimensional case for definiteness, we consider
${\Bbb S}^4_{sv}\simeq B^4\cup_{S^3}C(\partial{B^4})$, ({\it sv} for
singular vertex).
Note that any
triangulation of ${\Bbb S}^4$  can be factorized in this way
(since $C(\partial{B^4})$ and $\partial{B^4}$ are the star and the
link of a vertex, respectively), and we have
\begin{eqnarray}
N_4=N_4(B^4)+N_4(C(\partial{B^4})).
\end{eqnarray}

The triangulation is singular as soon as we have
\begin{eqnarray}
N_4(B^4)\propto {N}_4(C(\partial{B^4})),
\end{eqnarray}
namely when the cone $C(\partial{B^4})$  contains a number of top-
dimensional simplices growing linearly with the volume
of the whole manifold.

It is easily checked that the $f$-vector of ${\Bbb S}^4_{es}\simeq
B^4\cup_{S^3}C(\partial{B^4})$ is given by
\begin{eqnarray}
N_0&= & N_0(S^3)+ 1+ N_0(\hat{B}^4)\nonumber\\
N_1&= & N_1(S^3)+ N_0(S^3)+N_1(\hat{B}^4)\nonumber\\
N_2&= & N_2(S^3)+ N_1(S^3)+N_2(\hat{B}^4)\nonumber\\
N_3&= & N_3(S^3)+ N_2(S^3)+N_3(\hat{B}^4)\nonumber\\
N_4&= & N_3(S^3)+ N_4(B^4),
\label{esseffe}
\end{eqnarray}
where $N_i(S^3)$ denotes the $f$-vector of the boundary
$\partial(B^4)\simeq{\Bbb S}^3$ of the triangulated ball $B^4$, and
$N_i(\hat{B}^4)$ is the $f$-vector of the interior of $B^4$.
The Dehn-Sommerville relations for ${\Bbb S}^4$ and for
${\Bbb S}^3$ constrain $N_i(\hat{B}^4)$ and $N_k(S^3)$ according to
\begin{eqnarray}
& & N_0(\hat{B}^4)-N_1(\hat{B}^4)+N_2(\hat{B}^4)-N_3(\hat{B}^4)
+N_4({B}^4)=1\nonumber\\
& & 2N_1(\hat{B}^4)-3N_2(\hat{B}^4)+4N_3(\hat{B}^4)-
5N_4({B}^4)+ N_0(S^3)=0\nonumber\\
& & 2N_3(\hat{B}^4)+N_3(S^3)= 5N_4(\hat{B}^4).
\end{eqnarray}

The average incidence $b(4,2)$ of such triangulated ${\Bbb S}^4$ can
be easily computed in terms of the $f$-vectors $N_i(\hat{B}^4)$ and
$N_k(S^3)$ according to
\begin{eqnarray}
b(4,2)=10\cdot\frac{4b(3,1)N_3(S^3)+2b(3,1)[N_0(\hat{B}^4)-
N_1(\hat{B}^4)+N_2(\hat{B}^4)]-
2b(3,1)}{[6b(3,1)+18]N_3(S^3)+3b(3,1)N_2(\hat{B}^4)},
\label{deb}
\end{eqnarray}
where $b(3,1)\doteq6[N_3(S^3)/N_1(S^3)]$ is the average incidence of
$\partial{B^4}\simeq{\Bbb S}^3$. The presence of a singular vertex
corresponds to
\begin{eqnarray}
\frac{N_4(\hat{B}^4)}{N_3(S^3)}=O(1),
\end{eqnarray}
and it is easily verified that under such condition $b(4,2)$ is an
increasing function of $b(3,1)$. {\it This remark implies that
singular
triangulations with the smallest possible $b(4,2)$ are to be found
corresponding to $b(3,1)=b(3,1)_{min}=9/2$}.

We have already seen an example of such a triangulation in the
previous section, one for which the lowest kinematically possible
incidence, $b(2,4)=4$, is attained. However such examples are not
generic. They correspond to assuming $y_2=0$, (or more generally, they
still occur if one interprets the right hand side as $y_2=O(1)$), and
the
singular vertex is not stable under $(1,5)$ moves. Eventually by
performing enough barycentric subdivisions the initial singular vertex
is smoothed out. Explicitly,
assume that we start our  chain of barycentric subdivisions on an
a ${\Bbb S}^4_{sv}\simeq B^4\cup_{S^3}C(\partial{B^4})$ with a given
value of $N_4$, say $N_4(0)$. Denote by ${\Bbb S}^4_{sv}(0)$ this
initial
triangulation. Note that at this initial step
\begin{eqnarray}
N_4(B^4(0))=\frac{1}{3}{N}_4(C(\partial{B^4(0)})),
\end{eqnarray}
(see the previous section). If we carry out a $(1,5)$ move on each
$4$-simplex of ${\Bbb S}^4_{sv}(0)$, we get a triangulation of
${\Bbb S}^4$ still of the form ${\Bbb S}^4_{sv}\simeq
B^4\cup_{S^3}C(\partial{B^4})$, which we denote
${\Bbb S}^4_{sv}(1)$. For such triangulation we have
\begin{eqnarray}
{N}_4(C(\partial{B^4(1)}))= 4\cdot{N}_4(C(\partial{B^4(0)})),
\end{eqnarray}
\begin{eqnarray}
N_4(B^4(1))= 5\cdot{N}_4(B^4(0))+{N}_4(C(\partial{B^4(0)})).
\end{eqnarray}
Now proceed by induction, noticing that if at each step we carry out a
barycentric subdivision of each $4$-simplex of the ${\Bbb S}^4_{sv}$
generated at the previous step, we still get a $4$-sphere, ${\Bbb
S}^4_{sv}(k)$, triangulated according to
${\Bbb S}^4_{sv}\simeq B^4\cup_{S^3}C(\partial{B^4})$, and such that,
at
the $k$-th step,
\begin{eqnarray}
{N}_4(C(\partial{B^4(k)}))= 4^k\cdot{N}_4(C(\partial{B^4(0)})),
\end{eqnarray}
\begin{eqnarray}
N_4(B^4(k))=
5^k\cdot{N}_4(B^4(0))+{N}_4(C(\partial{B^4(0)}))\sum_{i=1}^{k-1}5^{i-
1}\cdot4^{k-1}.
\end{eqnarray}
Thus, as $k$ grows (corresponding to $y_1\to+\infty$, $y_2=O(1)$),
$N_4(B^4(k))$
largely dominates over
${N}_4(C(\partial{B^4(k)}))$:
\begin{eqnarray}
N_4(B^4(k))>>_{y_1\to+\infty}
{N}_4(C(\partial{B^4(k)})),
\end{eqnarray}
and the resulting triangulation of ${\Bbb S}^4$ is no longer singular.
In this sense, the dominance of the $(1,5)$ move naturally yields
regular stacked sphere and thus for branched polymers.
\vskip 0.5 cm
Discarding these particular examples of unstable triangulated spheres
with a singular vertex,
we can easily characterize the smallest $b(4,2)$ corresponding to
generic singular triangulations, namely triangulations generated in
the large volume limit as $y_1\to\infty$ and $y_2\to\infty$, and whose
singular vertices are stable under the action of the $(k,l)$ moves (at
a fixed ratio $\frac{y_1}{y_2}$).
Let us start by noticing that corresponding to $b(3,1)=9/2$, the
expression (\ref{deb}) for the average incidence reduces to
\begin{eqnarray}
b(4,2)=10\cdot\frac{6N_1(S^3)+4[N_0(\hat{B}^4)-
N_1(\hat{B}^4)+N_2(\hat{B}^4)]}{15N_1(S^3)+6N_2(\hat{B}^4)},
\label{qui}
\end{eqnarray}
where, in the numerator, we have discarded terms which are $o(1)$,
thus irrelevant in
the large volume limit.
\vskip 0.5 cm
Since ${\Bbb S}^3$ is {\it stacked}, the integers $N_3(S^3)$ and
$N_1(S^3)$ are related by $4N_3(S^3)=3N_1(S^3)-10$, which implies that
$3N_1(S^3)\equiv10(mod\;4)$. Thus, $N_1(S^3)$ must be an integer
multiple of $4$ up to an error term which goes to zero, with
increasing $N_1(S^3)$, as $10/N_1(S^3)$.
More explicitly, and referring to the expression of the $f$-vector of
${\Bbb S}^3$ in terms of the balance variables $x_1\in{\Bbb N}$ and
$x_2\in{\Bbb N}$ introduced in section \ref{ergodic}, we get
the following components:
\begin{eqnarray}
N_0(S^3)&=&5+x_1,\nonumber\\
N_1(S^3)&=&10+4x_1,\nonumber\\
N_2(S^3)&=&10+6x_1,\nonumber\\
N_3(S^3)&=&5+3x_1,
\end{eqnarray}
since corresponding to a stacked ${\Bbb S}^3$ we have $x_2=0$, (see
(\ref{tredim})).
\vskip 0.5 cm
The congruence properties just established for the $f$-vector of a
stacked $3$-sphere suggest
to parameterize  both $N_2(\hat{B}^4)$ and   $N_0(\hat{B}^4)-
N_1(\hat{B}^4)+N_2(\hat{B}^4)$, appearing in (\ref{qui}), in terms of
$N_1(S^3)$ by setting
\begin{eqnarray}
N_2(\hat{B}^4)=\tilde{\beta}N_1(S^3),
\end{eqnarray}
and
\begin{eqnarray}
N_0(\hat{B}^4)-N_1(\hat{B}^4)+N_2(\hat{B}^4)]=\tilde{\alpha}N_1(S^3).
\end{eqnarray}
According to the above remarks, $N_1(S^3)$ is asymptotically an
integer multiple of $4$, thus if we are interested to triangulations
for which $N_1(S^3)$ can grow arbitrarily large, it follows that
the two parameters $\tilde{\alpha}$ and $\tilde{\beta}$
necessarily are rational numbers of the form
$\tilde{\beta}=\frac{\beta}{4}$ and
$\tilde{\alpha}=\frac{\alpha}{4}$ with $\beta$ and $\alpha$ integers.
In other words,
the generic triangulations of
${\Bbb S}^4_{sv}\simeq B^4\cup_{S^3}C(\partial{B^4})$ with the joining
${\Bbb S}^3$ stacked ($b(3,1)=9/2$), can be conveniently parameterized
by setting
\begin{eqnarray}
\frac{N_2(\hat{B}^4)}{N_1(S^3)}\doteq\frac{\beta}{4},
\label{paruno}
\end{eqnarray}
and
\begin{eqnarray}
\frac{N_0(\hat{B}^4)-
N_1(\hat{B}^4)+N_2(\hat{B}^4)}{N_1(S^3)}\doteq\frac{\alpha}{4},
\label{pardue}
\end{eqnarray}
where  $\alpha$ and $\beta$ are integers. Note that while
$\beta\geq0$,  $\alpha$ can possibly take also negative values.
However, if we rewrite (\ref{qui}) in terms of such parameters
\begin{eqnarray}
b(4,2)=10\cdot\frac{12+2\alpha}{30+3\beta},
\label{crumpledb}
\end{eqnarray}
the kinematical bound $b(4,2)\geq4$ implies $5\alpha\geq3\beta$, and
thus $\alpha$ is non-negative as well.
\vskip 0.5 cm

The
parameters $\alpha$ and $\beta$ so introduced are completely
equivalent to the balance variables $y_1$ and $y_2$ related to the
cumulant action of the $(k,l)$ moves. Explicitly, we obtain
\begin{eqnarray}
2y_1 &= & (\frac{1}{2}+\frac{1}{4}\beta-\frac{1}{3}\alpha)N_1(S^3)-
\frac{20}{3}\nonumber\\
2y_2 &= & (\frac{5}{6}\alpha-\frac{1}{2}\beta)N_1(S^3)+\frac{20}{3}
\label{ipsilon}.
\end{eqnarray}
The Dehn-Sommerville relations for the $f$-vector $N_i(\hat{B}^4)$
allow us to express also its components in terms of $\alpha$ and
$\beta$ according to
\begin{eqnarray}
3N_0(\hat{B}^4) &= & \left[\frac{3\beta-
4\alpha}{8}\right]{N}_1(S^3)+10\nonumber\\
3N_1(\hat{B}^4) &= & \left[\frac{9\beta-
10\alpha}{8}\right]{N}_1(S^3)+10\nonumber\\
N_2(\hat{B}^4) &= & \frac{1}{4}\beta{N}_1(S^3)\nonumber\\
3N_3(\hat{B}^4) &= & \left[\frac{3+5\alpha}{4}\right]{N}_1(S^3)-
5\nonumber\\
3N_4(\hat{B}^4) &= & \left[\frac{3+2\alpha}{4}\right]{N}_1(S^3)-2.
\label{dellapalla}
\end{eqnarray}

The generic conditions $y_1>0$ and $y_2>0$, (and both
approaching $+\infty$), together with $N_0(\hat{B}^4)>0$,
imply that the parameters $\alpha$ and
$\beta$ are related by
\begin{eqnarray}
\frac{3}{5}\beta<\alpha<\frac{3}{4}\beta,
\label{alfabeta}
\end{eqnarray}
with $(\alpha,\beta)\in{\Bbb N}^+\times{\Bbb N}^+$. From these remarks
it follows that, as $y_1$ and $y_2$ go to $+\infty$, there are  two
distinct regimes for the set of triangulations considered:
\vskip 0.3 cm

\noindent {\it (i)} If we {\it constrain} the $f$-vector $N_i(S^3)$ of
the connecting
$\partial{B}^4$ to be $O(1)$, then  according to (\ref{ipsilon}),
$\alpha$ and $\beta$ go to $\infty$  as $y_1,y_2\to+\infty$.
From (\ref{dellapalla}) we get that in this regime
\begin{eqnarray}
N_4(S^4) &\simeq & N_4(B^4)\simeq(\alpha/6)N_1(S^3)\nonumber\\
N_2(S^4) &\simeq & N_2(B^4)=(\beta/4){N_1}(S^3),
\end{eqnarray}
where $N_1(S^3)$ is a constant. The geometrical  bounds
(\ref{alfabeta}) simply imply that as $\alpha,\beta\to\infty$, the
corresponding average incidence $b(4,2)$ varies between the
kinematical bounds $4\leq b(4,2)\leq5$, as required.
\vskip 0.3 cm

\noindent {\it (ii)} Conversely, if we do not constrain $N_i(S^3)$ to
be $O(1)$, then
according to (\ref{ipsilon}), $N_1(S^3)$, (and hence $N_3(S^3)$), is
allowed to grow unboundedly large as $y_1,y_2\to+\infty$. This growth,
which corresponds to the generation of singular vertices, is possible
for any finite value of the parameters $\alpha$ and $\beta$ compatible
with (\ref{alfabeta}). Note that if kinematically possible, according
to (\ref{alfabeta}), such singular triangulations {\it entropically
dominate} over the regular ones since these latter are generated by
the constrained configurations forcing $N_i(S^3)$ to be $O(1)$, while
the former are unconstrained. More specifically, since the number of
distinct triangulations of a $3$-sphere ${\Bbb S^3}$ grows
exponentially with $N_3(S^3)$, configurations with $N_3(S^3)$ as large
as possible, if kinematically allowed, will dominate over
configurations with $N_3(S^3)=O(1)$.
\vskip 0.5 cm

The kinematical bound (\ref{alfabeta}) for the occurrence of singular
triangulations is not trivial. In order to discuss its implications,
let us consider the ratio
between the total volume of the triangulated ${\Bbb S}^4_{sv}$ and the
volume of the ball around the singular vertex $\sigma^0$, {\it viz.}
\begin{eqnarray}
\frac{Vol(S^4)}{Vol_{sing}(\sigma^0)}=\frac{N_4}{N_3(S^3)}=
\frac{12+2\alpha}{9}.
\label{volsing}
\end{eqnarray}
A direct computation of
the average incidence $b(4,2)$, (see \ref{crumpledb}), together with
(\ref{alfabeta}) immediately shows that the {\it
smallest} $b(4,2)$'s for which we may have singular triangulations
occur for
\begin{eqnarray}
\alpha &= & 5+3h,\nonumber\\
\beta &= & 8+5h,
\end{eqnarray}
with $h=0,1,2,\ldots$. As $h$ varies, the average incidence $b(4,2)_h$
and the volume ratio (\ref{volsing}) respectively take the values:
\begin{eqnarray}
b_h(4,2)=10\cdot\frac{22+6h}{54+15h},
\label{criticalbs}
\end{eqnarray}
\begin{eqnarray}
\frac{Vol(S^4)}{Vol_{sing}(\sigma^0)}\bigg|_{h}=\frac{22+6h}{9}.
\label{critvol}
\end{eqnarray}
Since the singular triangulations that entropically dominate are those
for which
$\frac{Vol(S^4)}{Vol_{sing}(\sigma^0)}\Big|_{h}$ is as low as
possible,
the smallest $b(4,2)$
for which we may have {\it generic} singular triangulations with the
largest $Vol_{sing}(\sigma^0)$ is
\begin{eqnarray}
b(2,4)_{sing}=\frac{110}{27}\simeq 4.07407\ldots,
\end{eqnarray}
(corresponding to $h=0$ and
$\frac{Vol(S^4)}{Vol_{sing}(\sigma^0)}=22/9\simeq2.444\ldots$).
It is easily verified that such an average incidence is associated to
a relative concentration of $(1,5)$ moves versus $(2,4)$ moves given
by
\begin{equation}
y_1=5y_2.
\label{concentration}
\end{equation}
\vskip 0.5 cm
It is clear from (\ref{criticalbs}) that singular triangulations may
appear also for smaller values of $b(4,2)$. A
list of the first possible values of $b_h(4,2)$ is provided by table
\ref{tavola1}. These triangulations are
less singular than the ones associated with
$b(4,2)=\frac{110}{27}$ since they correspond to larger values of the
ratio $\frac{Vol(S^4)}{Vol_{sing}(\sigma^0)}$, and for this reason we
may be
tempted to consider them as
entropically {\it sub dominating} at least in the large-volume limit.
Yet, this is mere
apparency since their presence is
particularly relevant for locating the critical incidence $b_0$ and
for understanding the present status of the Monte Carlo simulations.
\begin{table}[t]
\begin{center}
\begin{tabular}{|c||c||c||c||c|}
h & $\beta$ & $\alpha$ & b(2,4) &
$\frac{Vol(S^4)}{Vol_{sing}(\sigma^0)}|_h$ \\ \hline
0 & 8 & 5 & $\frac{110}{27}\simeq4.07407$ & $\frac{22}{9}\simeq2.444$
\\
1 & 13 & 8 & $\frac{280}{69}\simeq4.0579$ & $\frac{28}{9}\simeq3.111$
\\
2 & 18 & 11 & $\frac{340}{84}\simeq4.04761$ &
$\frac{34}{9}\simeq3.777$ \\
3 & 23 & 14 & $\frac{400}{99}\simeq4.0404$ & $\frac{40}{9}\simeq4.444$
\\
4 & 28 & 17 & $\frac{460}{114}\simeq4.03508$ &
$\frac{46}{9}\simeq5.111$ \\
5 & 33 & 20 & $\frac{520}{129}\simeq4.03100$ &
$\frac{52}{9}\simeq5.777$ \\
\end{tabular}
\end{center}
\vskip .05 cm
\caption{The smallest incidence numbers $b(4,2)|_h$ and the associated
singular volume fraction
$Vol(S^4)/Vol_{sing}(\sigma^0)$ as a function of the parameters
$\alpha$ and $\beta$.}
\label{tavola1}
\end{table}
Moreover, as we see in the next section, these triangulations have a
subtle interplay with
the particular singular geometry dominating in the strong coupling
phase of $4$-D simplicial gravity: PL-manifolds with a single singular
edge connecting two singular vertices.
\vskip 0.5 cm
In order to get the complete geometrical picture, one has to note
that for $\alpha=2+8h$ and  $\beta=3+13h$, we also get a highly
degenerate configuration for which
\begin{equation}
b_h(4,2)=\frac{160}{39}\simeq 4.102564
\end{equation}
is a constant average incidence as $h$ varies, whereas
$\frac{Vol(S^4)}{Vol_{sing}(\sigma^0)}|_{h}=\frac{16+16h}{9}$,
$h=0,1,2,\ldots$.

In other words, corresponding to such value of $b(4,2)$ we have
distinct triangulations
with distinct ratios $\frac{Vol(S^4)}{Vol_{sing}(\sigma^0)}|_{h}$ but
with $b(4,2)$ fixed.
Even if this set of triangulations contains configurations for which
$\frac{Vol(S^4)}{Vol_{sing}(\sigma^0)}|_{h}\simeq1.777$, such a
degeneration
makes any particular configuration at fixed
$\frac{Vol(S^4)}{Vol_{sing}(\sigma^0)}|_{h}$ entropically sub-
dominating
with respect to the generic configurations described by
(\ref{criticalbs}), at least as $N_4\to\infty$.

\subsection{The development of singular edges: ${\Bbb S}^4_{es}$}
\label{sedge}
The explicit construction of the previous section may suggest that the
singular triangulations we are explicitly considering are
characterized by the dominance of just one singular vertex. Actually,
as the parameters $\alpha$ and $\beta$ vary, triangulations of
${\Bbb S}^4$ of the form $B^4\cup_{S^3}C(\partial{B^4})$ are not the
only ones possible whose average incidence $b(4,2)$ takes on the value
(\ref{crumpledb}), {\it at least as} $N_4(S^4)\to\infty$. As a matter
of fact, in the infinite volume limit, (but not at finite volume),
triangulations,
with more than one singular vertex and with singular edges are still
characterized by
the average incidence (\ref{crumpledb}). Their dominance in the class
of triangulations considered, as $N_4(S^4)\to\infty$, is driven by a
rather simple entropic mechanism which we discuss in detail in this
section.
\vskip 0.5 cm

Implicitly, the occurrence of more than one singular vertices may
still be described by the construction
$B^4\cup_{S^3}C(\partial{B^4})$, since one may simply consider the new
singular vertices and edges to be located in the ball $B^4$. However,
the interplay between dominance of one or more (edge-connected)
singular vertices is most easily seen from a simple variant of the
construction leading to (\ref{crumpledb}). The generic singular
triangulation of ${\Bbb S}^4$ is still realized by glueing two $4$-
balls along an isometric ${\Bbb S}^3$ boundary which is again assumed
to be a
stacked $3$-sphere, {\it i.e.}, as $B_{es}^4\cup_{S^3}{B^4}$. However,
one of the two balls, say the one denoted by $B_{es}^4$, ($es$ being
an acronym for {\it edge-singular}), is no longer taken of the form of
a cone $C(\partial{B^4})$ over the ${\Bbb S}^3$ boundary, but more
generally is provided by a triangulation with $f$-vector
\begin{eqnarray}
N_0(B^4_{es})&=&\frac{1}{3}\sum_{j=1}^kN_3(B^3(j))+k\nonumber\\
N_1(B^4_{es})&=&\frac{5}{3}\sum_{j=1}^kN_3(B^3(j))+\frac{1}{2}\sum_{l=
1}^{k-1}N_2(S^2(l))+
3(k-1)\nonumber\\
N_2(B^4_{es})&=&\frac{10}{3}\sum_{j=1}^kN_3(B^3(j))+2\sum_{l=1}^{k-
1}N_2(S^2(l))+
2(k-1)\nonumber\\
N_3(B^4_{es})&=&3\sum_{j=1}^kN_3(B^3(j))+\frac{5}{2}\sum_{l=1}^{k-
1}N_2(S^2(l))\nonumber\\
N_4(B^4_{es})&=&\sum_{j=1}^kN_3(B^3(j))+\sum_{l=1}^{k-1}N_2(S^2(l)).
\label{peanut}
\end{eqnarray}
To grasp the geometrical origin of this $f$-vector imagine $k$
distinct $3$-spherical disks,
$B^3(j)$, joined through $(k-1)$ ${\Bbb S}^2$-boundaries, ${\Bbb
S}^2(l)$; a sort of $3$-dimensional {\it peanut-shell} with $k$-bulges
and $k-1$ necks. This gives rise to a $3$-spherical peanut-shell
${\Bbb S}^3$, and we get a $4$-dimensional ball $B^4_{es}$ out of this
${\Bbb S}^3$ by considering $(k-1)$-edges
$\{\sigma^1(l)\}_{l=1,\ldots,k-1}$ not belonging to
${\Bbb S}^3$ and connecting $k$ vertices
$\{\sigma^0(j)\}_{j=1,\ldots,k}\notin{\Bbb S}^3$.
The $4$-ball $B^4_{es}$ is defined by requiring that the generic $2$-
spheres ${\Bbb S}^2(l)$ are the links (in $B^4_{es}$) of the
corresponding edges $\sigma^1(l)$, $l=1,\ldots,k-1$. Moreover, the
complex obtained by $B^4_{es}$ by removing the $(k-1)$ stars (in
$B^4_{es}$) of the edges
$\sigma^1(l)$, is assumed to be the disjoint union of $k$ cones
$C(B^3(j))$ over the $3$-spherical disks $B^3(j)$, with apices in the
$k$ vertices $\sigma^0(j)$. This construction can be roughly
described as a $3$-dimensional peanut shell containing one rather than
$k$ distinct $4$-dimensional nuts. It is easily verified that the $f$-
vector $N_i(B^4_{es})$ (\ref{peanut})
describes this generalized {\it peanut triangulation} and that it
represents
a triangulated $4$-ball with $N_0(B^4_{es})$ vertices, $k$ of which,
$\{\sigma^0(j)\}_{j=1,\ldots,k}$ are interior vertices, ({\it i.e.},
$\sigma^0(j)\notin\partial{B^4_{es}}$), with $N_3(B^3(j))+N_2(S^2(j))$
$4$-simplices $\sigma^4$ incident on the $j$-th of them. The  $j$-th
of the $k-1$
interior links $\sigma^1(l)$, connects the vertex $\sigma^0(j)$ with
$\sigma^0(j+1)$, and
$N_2(S^2(j))$ $4$-dimensional simplices $\sigma^4$ are incident on it.
Thus, if
some, say $1\leq{s}\leq{k}$, of the $\{N_3(B^3(j)\}$ and the
corresponding $s-1$ of the $\{N_2(S^2(l))\}$ grow with the simplicial
volume of the ${\Bbb S}^4\supset{B^4_{es}}$,
(not necessarily with the same rate), the
triangulation of $B^4_{es}$ just constructed contains $s$ {\it
singular} vertices connected by
$s-1$ {\it singular} edges. Note that if we take the boundary of this
triangulated $B^4_{es}$ we obtain a stacked $3$-sphere ${\Bbb S}^3$
with $f$-vector
\begin{eqnarray}
N_0(S^3)&=&\frac{1}{3}\sum_{j=1}^kN_3(B^3(j))\nonumber\\
N_1(S^3)&=&\frac{4}{3}\sum_{j=1}^kN_3(B^3(j))\nonumber\\
N_2(S^3)&=&2\sum_{j=1}^kN_3(B^3(j))\nonumber\\
N_3(S^3)&=&\sum_{j=1}^kN_3(B^3(j)).
\end{eqnarray}
This ${\Bbb S}^3$ boundary of the $4$-dimensional ball $B^4_{es}$ may
be profitably thought of as resulting from the connected sum, along
isometric ${\Bbb S}^2$-boundaries of $k$ distinct stacked $3$-spheres
${\Bbb S}^3_i$, $i=1,\ldots{k}$, to be considered as the links (in an
${\Bbb S}^4$) of a corresponding singular vertex. In this way the
singular ball $B^4_{es}$, (and the corresponding ${\Bbb S}^4$), can be
considered as the kinematical set up for discussing the {\it
interaction} of $k$ distinct singular vertices (of the type considered
in the previous section). This picture allow also to prove an
elementary but important result showing that, in the class of
triangulations considered, the order of singularity of a singular edge
$\sigma^1(l)$ in $B^4_{es}$ is subdominating with respect to the order
of singularity of the corresponding vertices $\sigma^0(l-1)$ and
$\sigma^0(l)$. In other words, in the large volume limit, the number
of $4$-simplices incident on $\sigma^1(l)$ grows slower than the
number of $4$-simplices incident on the vertices $\sigma^0(l-1)$ and
$\sigma^0(l)$. As usual, the number of incident $4$-simplices can be
considered as the (possibly) singular volume associated to the
corresponding edge or vertex. Thus, if we denote such simplicial
volumes by
$Vol(\sigma^1(l))=\#\{\sigma^4\cap\sigma^1(l)\}$,
$Vol(\sigma^0(l-1))=\#\{\sigma^4\cap\sigma^0(l-1)\}$, and
$Vol(\sigma^0(l))=\#\{\sigma^4\cap\sigma^0(l)\}$, we have the
following

\begin{lemma}
In the class of triangulations considered for $B^4_{es}$,
\begin{eqnarray}
\lim_{N_4(B^4_{es})\to\infty}\frac{Vol(\sigma^1(l))}{Vol(\sigma^0(l))}
=0.
\end{eqnarray}
\label{subessedue}
\end{lemma}
(Obviously the same holds with $Vol(\sigma^0(l))$ replaced by
$Vol(\sigma^0(l-1))$).
In order to prove this result we may consider, without loss in
generality, a $B^4_{es}$ whose ${\Bbb S}^3$-boundary consists of two
stacked $3$-spheres ${\Bbb S}_1$ and ${\Bbb S}_2$ joined through an
isometric ${\Bbb S}^2$. The more general case can be proved similarly
without much effort. Again without loss of generality we may assume
that the two isometric copies of ${\Bbb S}^2$ along which the two
${\Bbb S}^3_1$ and ${\Bbb S}^3_2$ are glued are the links of a vertex
in the corresponding ${\Bbb S}^3_i$, (as remarked in the previous
paragraph, this can be always arranged; also it can be easily shown
that there are stacked $3$-spheres with a marked vertex whose $2$-
spherical link grows linearly with the simplicial volume of the $3$-
sphere-see
section \ref{sss} for an example in $4$-dimensions. The above lemma
states that, in the case of stacked $3$-spheres, this large volume
behavior for an ${\Bbb S}^2$ cannot hold if such a $2$-sphere is a
joining neck). It immediately follows that the $f$-vector of
$\partial{B^4_{es}}={\Bbb S}^3_1\cup_{S^2}{\Bbb S}^3_2$ can be written
in terms of the $f$-vector of ${\Bbb S}^3_i$, $i=1,2$, and of ${\Bbb
S}^2$ as

\begin{eqnarray}
N_0(S^3)&=&\frac{1}{3}[N_3(S^3_1)+N_3(S^3_2)]-N_2(S^2)+N_0(S^2)-
6\nonumber\\
N_1(S^3)&=&\frac{4}{3}[N_3(S^3_1)+N_3(S^3_2)]-4N_2(S^2)+N_1(S^2)-
4\nonumber\\
N_2(S^3)&=&2[N_3(S^3_1)+N_3(S^3_2)]-4N_2(S^2)\nonumber\\
N_3(S^3)&=&N_3(S^3_1)+N_3(S^3_2)-2N_2(S^2).
\label{essetre}
\end{eqnarray}
Since $N_1(S^2)=(3/2)N_2(S^2)$ we immediately get
\begin{eqnarray}
\frac{N_3(S^3)}{N_1(S^3)}=\frac{3}{4}\cdot\frac{N_3(S^3_1)+N_3(S^3_2)-
2N_2(S^2)}{N_3(S^3_1)+N_3(S^3_2)-\frac{15}{8}N_2(S^2)-\frac{12}{4}},
\end{eqnarray}
which implies that $\frac{N_3(S^3)}{N_1(S^3)}>\frac{3}{4}$ as long as
$\frac{N_2(S^2)}{N_3(S^3_i)}=O(1)$ in the large volume limit ({\it
i.e.}, as
$N_3(S^3)+N_2(S^2)\to\infty$).
Thus $\partial{B^4_{es}}={\Bbb S}^3_1\cup_{S^2}{\Bbb S}^3_2$ can be a
stacked $3$-sphere
if and only if
\begin{eqnarray}
\lim_{N_3(S^3)\to\infty}\frac{N_2(S^2)}{N_3(S^3_i)}=0.
\label{subdom}
\end{eqnarray}
According to the above remarks $N_2(S^2)=Vol(\sigma^1(i))$ and
$N_3(S^3_i)+N_2(S^2)=Vol(\sigma^0(i))$, and we can write
(\ref{subdom}) as
\begin{eqnarray}
\lim_{N_4(B^4_{es})\to\infty}\frac{Vol(\sigma^1(i))}{Vol(\sigma^0(i))-
Vol(\sigma^1(i))}=0,
\end{eqnarray}
from which the lemma follows.
\vskip 0.5 cm
This latter result only implies that the singular volume of the edge
cannot grow {\it linearly} with the total volume of the ball
$B^4_{es}$ (and of the resulting
${\Bbb S}^4$-see below). As we have seen in the previous section, a
linear growth is instead typical for the singular volume associated to
the vertices. It should be stressed that a subdominant rate of growth,
(say with some fractional power of the total volume of
$B^4_{es}$), is well in agreement with (\ref{subdom}). As a matter of
fact subdominant powers for the volume growth associated with a
singular edge are the ones typically experienced in numerical
simulations \cite{Catterall}.
\vskip 0.5 cm
Note that triangulations with $Vol(\sigma^1(i))$ as large as
kinematically possible, thus growing with $[N_4(B^4_{es})]^{\delta}$
for some $0<\delta<1$, entropically dominate over triangulations of
$B^4_{es}$ with $Vol(\sigma^1(i))=O(1)$. This remark follows as a
direct consequence of the fact that triangulating $B^4_{es}$ under the
hypothesis $N_2(S^2)=O(1)$, while sufficient to assure the validity of
(\ref{subdom}), it is not a necessary condition. It generates a
subclass of constrained configurations in the class of triangulations
of $B^4_{es}$ considered. Conversely, triangulations with
$N_2(S^2)\propto[N_4(B^4_{es})]^{\delta}$ are, according to
(\ref{subdom}), unconstrained, and as such much more numerous at least
in the large volume limit.
\vskip 0.5 cm
As in the previous section, we obtain a $4$-sphere ${\Bbb S}^4_{es}$,
({\it es} again for {\it edge-singular}),  by glueing a generic
triangulated ball $B^4$ with stacked ${\Bbb S}^3$ boundary to the
singular $B^4_{es}$ defined by (\ref{peanut}),
{\it viz.}, ${\Bbb S}^4_{es}\simeq{B^4}\cup_{S^3}B^4_{es}$. It is
easily checked that the $f$-vector of such an ${\Bbb S}^4_{es}$ is
given by
\begin{eqnarray}
N_0(S^4_{es})&= & N_0(S^3)+ N_0(\hat{B}^4)+k\nonumber\\
N_1(S^4_{es})&= & N_1(S^3)+ N_0(S^3)+N_1(\hat{B}^4)+
\frac{1}{2}\sum_{l=1}^{k-1}N_2(S^2(l))+3(k-1)\nonumber\\
N_2(S^4_{es})&= & N_2(S^3)+ N_1(S^3)+N_2(\hat{B}^4)+
2\sum_{l=1}^{k-1}N_2(S^2(l))+2(k-1)\nonumber\\
N_3(S^4_{es})&= & N_3(S^3)+ N_2(S^3)+N_3(\hat{B}^4)+
\frac{5}{2}\sum_{l=1}^{k-1}N_2(S^2(l))\nonumber\\
N_4(S^4_{es})&= & N_3(S^3)+ N_4(B^4)+\sum_{l=1}^{k-1}N_2(S^2(l)),
\end{eqnarray}
where $N_i(S^3)$ denotes the $f$-vector of the joining stacked $3$-
sphere
$\partial(B^4_{es})\simeq{\Bbb S}^3$, and
$N_i(\hat{B}^4)$ is the $f$-vector of the interior of $B^4$. According
to (\ref{subdom}),
$N_2(S^2)/N_3(S^3)$ is asymptotically $o(1)$, thus, in the large
volume limit, the average incidence $b(4,2)$ of such a triangulated
${\Bbb S}^4_{es}$ is still provided by the expression
(\ref{crumpledb}) introduced in the previous section, {\it viz.},
\begin{eqnarray}
\lim_{N_4(S^4)\to\infty}b(4,2)|_{S^4_{es}}=10\cdot\frac{12+2\alpha}{30
+3\beta},
\label{crumpledb2}
\end{eqnarray}
where the two parameters $\beta$ and $\alpha$ are again defined by
(\ref{paruno}) and (\ref{pardue}), respectively. Before we proceed any
further, we should emphasize that
(\ref{crumpledb2}) strictly speaking only holds in the limit
$N_4(S^4)\to\infty$, and that {\it at finite} (but large) volume
$N_4(S^4_{es})$, we have $b(4,2)|_{S^4_{es}}>b(4,2)$ with
\begin{eqnarray}
b(4,2)|_{S^4_{es}}=10\cdot\frac{12+2\alpha}{30+3\beta}+\eta\frac{\sum_
{l=1}^{k-1}N_2(S^2(l))}{N_3(S^3)},
\label{crumpledb3}
\end{eqnarray}
for a suitable $\alpha$- and $\beta$-dependent constant $\eta>0$ which
can be easily worked out.
For instance, for the relevant case $k=2$, ({\it i.e.}, two singular
vertex connected by a sub-singular edge), we get to leading order in
$N_2(S^2)/N_3(S^3)$,
\begin{eqnarray}
b(4,2)|_{S^4_{es}}=10\cdot\frac{12+2\alpha}{30+3\beta}+
10\cdot\frac{6+3\beta-4\alpha}{100+\beta^2+20\beta}
\left[\frac{N_2(S^2)}{N_3(S^3)}\right].
\label{bexample}
\end{eqnarray}
Since according to
lemma \ref{subessedue} the ratio $\frac{\sum_{l=1}^{k-
1}N_2(S^2(l))}{N_3(S^3)}$  can go to zero, in the large volume limit,
as slowly as $N_3(S^3)^{\delta-1}$ for some $0<\delta<1$, we get the
\begin{lemma}
At finite volume $N_4(S^4)$, the singular-vertex triangulations ${\Bbb
S}^4_{sv}\simeq{B^4}\cup_{S^3}C(\partial{B^4})$, considered in section
\ref{singvertex}, are closer to the kinematical boundary $b(4,2)=4$
than the edge-singular triangulations
${\Bbb S}^4_{es}\simeq{B^4}\cup_{S^3}B^4_{es}$.
\label{closer}
\end{lemma}

We stress that this result does not imply that  the singular-vertex
triangulations
${\Bbb S}^4_{sv}\simeq{B^4}\cup_{S^3}C(\partial{B^4})$ entropically
dominate in the large volume limit. For, according to
(\ref{crumpledb2}), the edge-singular triangulations
become more and more important as the volume increases,
and eventually in the infinite volume limit the triangulated spheres
${\Bbb S}^4_{es}$ enter in full entropic competition with the
triangulated ${\Bbb S}^4_{sv}$ considered in the previous section.
Actually this entropic competition comes into play quite rapidly as
the volume increases.
For instance, from (\ref{bexample}), one gets that, for the dominating
configurations at $h=0$,
\begin{eqnarray}
b(4,2)|_{S^4_{es}}\simeq \frac{110}{27}+\frac{100}{324}\cdot
\left[\frac{N_2(S^2)}{N_3(S^3)}\right].
\label{edgexample}
\end{eqnarray}
Numerical simulations at $N_4(S^4)=32000$, (see {\it e.g.},
\cite{Catterall} and
\cite{singedge}) show evidence that $N_2(S^2)/N_3(S^3)<1/10$, thus the
average incidence
$b(4,2)|_{S^4_{es}}$ of ${\Bbb S}^4_{es}$ differs (at $h=0$) from the
average incidence
$b(4,2)|_{S^4_{vs}}$ of ${\Bbb S}^4_{vs}$ by less than $3/100$.
Therefore it is important to understand how, as $N_4(S^4)$ increases,
the $k$ distinct singular vertices (and the corresponding $k-1$
subsingular connecting edges) interact among them, and which
configuration actually dominates in the large volume limit.
\vskip 0.5 cm
As we have seen in section \ref{singvertex}, the various singular
triangulations of the $4$-sphere
considered there are parameterized
by the ratio between the total simplicial volume of the given ${\Bbb
S}^4_{sv}$ and the simplicial volume of its singular part, (see
(\ref{volsing}) and (\ref{critvol})). If we consider a similar ratio
also for ${\Bbb S}^4_{es}$, {\it i.e.},
\begin{eqnarray}
\frac{Vol(S^4_{es})}{Vol(sing)}=
\frac{N_4(S^4_{es})}{N_4(B^4_{es})},
\label{sing3}
\end{eqnarray}
then, as is easily verified, this ratio is still provided, in the
large volume limit, by (\ref{volsing}). It follows that the entropic
comparison between the single singular vertex triangulations ${\Bbb
S}^4_{sv}$ and the multiple singular vertices triangulations ${\Bbb
S}^4_{es}$ should be carried out at a fixed value of the ratio
$Vol(S^4_{es})/Vol(sing)=const.=Vol(S^4_{sv})/Vol_{sing}(\sigma^0)$.

In our case
\begin{eqnarray}
N_4(B^4_{es})&=&\sum_{j=1}^kN_3(B^3(j))+\sum_{l=1}^{k-
1}N_2(S^2(l))\nonumber\\
&=&N_3(S^3)+\sum_{l=1}^{k-1}N_2(S^2(l)).
\label{ennesse}
\end{eqnarray}
According to the remarks following lemma \ref{subessedue},
unconstrained triangulations of ${\Bbb S}^4_{es}$ generally have
$\sum_{l=1}^{k-1}N_2(S^2(l))=O(N_3(S^3)^{\delta})$ for some
$0<\delta<1$.
Thus
\begin{eqnarray}
\lim_{N_4(S^4)\to\infty}N_4(B^4_{es})/N_3(S^3)=1,
\end{eqnarray}
and working at constant ratio (\ref{sing3}), (in the infinite volume
limit $N_4(S^4_{es})\to\infty$), implies that we have to consider
triangulations of $B^4_{es}$ with
\begin{eqnarray}
N_3(S^3)=A_1\cdot N_4(S^4)
\end{eqnarray}
and
\begin{eqnarray}
A_2\leq\sum_{l=1}^{k-1}N_2(S^2(l))\leq A_3\cdot N_3(S^3)^{\delta},
\end{eqnarray}
for some positive constants $A_1$, $A_2$, and $A_3$.
\vskip 0.5 cm

Guided by these considerations we can easily get a set of entropic
rules for determining which configuration dominates in the set of
singular triangulations of ${\Bbb S}^4$. We start
by an obvious adaptation of an argument in \cite{Catterall}, according
to which the number of distinct triangulations associated with a
singular vertex, (the {\it local entropy of the vertex}), is provided
by the number of distinct triangulations of the link of the given
vertex. The link, $link(\sigma^0(j))$, around the $j$-th singular
vertex $\sigma^0(j)\in{B^4_{es}}$, is a $3$-sphere ${\Bbb S}^3(j)$,
and any two such links,
${\Bbb S}^3(j)$ and ${\Bbb S}^3(j+1)$, associated with two singular
vertex connected by a singular edge $\sigma^1(j)$, have a non-empty
intersection ${\Bbb S}^2(j)$, (the link of the
connecting edge $\sigma^1(j)$). Thus, the inclusion-exclusion
principle implies that the number,
$Card[B^4_{es}(S^3(1),\ldots,S^3(k);S^2(1),\ldots,S^2(k-1))]$, of
distinct triangulations of $B^4_{es}$ with given singular vertices
$\{S^3(j)\}_{j=1,\ldots,k}$ and given singular edges
$\{S^2(l)\}_{l=1,\ldots,k-1}$ is provided by
\begin{eqnarray}
Card[B^4_{es}(S^3(1),\ldots,S^3(k);S^2(1),\ldots,S^2(k-
1)))]=\frac{\prod_{j=1}^kCard[S^3(j)]}{\prod_{l=1}^{k-1}Card[S^2(l)]},
\end{eqnarray}
where $Card[S^3(j)]$ and $Card[S^2(l)]$ respectively denote the number
of distinct triangulations of the $3$-spherical links of the $j$-th
singular vertex and of the $2$-spherical singular link of the $l$-th
singular edge. Since each $S^3(j)$ is a stacked $3$-sphere, (hence
with an average incidence $b(3,1)=\frac{9}{2}$), the microcanonical
partition function (\ref{asintotica}) immediately provides the leading
order asymptotics both for $Card[S^3(j)]$ and $Card[S^2(l)]$, {\it
viz.},
\begin{eqnarray}
Card[S^3(j)]_{N_3(S^3(j))>>1}\simeq
\left [
\frac{(b(3,1)-\hat{q}+1)^{b(3,1)-\hat{q}+1}}{(b(3,1)-\hat{q})^{b(3,1)-
\hat{q}}}
\right ] ^{N_1(S^3(j))},
\end{eqnarray}
where $b(3,1)=\frac{9}{2}$, $\hat{q}=3$. Since
$N_1(S^3(j))=\frac{4}{3}N_3(S^3(j))$, we get
\begin{eqnarray}
Card[S^3(j)]_{N_3(S^3(j))>>1}\simeq
\left [
\frac{(\frac{5}{2})^{5/2}}{(\frac{3}{2})^{3/2}}
\right ] ^{\frac{4}{3}N_3(S^3(j))}.
\end{eqnarray}
Similarly, by setting $b(2,1)=6$, $\hat{q}=3$, and
$N_0(S^2(l))=2+\frac{1}{2}N_2(S^2(l))$, (\ref{asintotica}) provides
\begin{eqnarray}
Card[S^2(l)]_{N_2(S^2(l))>>1}\simeq
\left [
\frac{(b(2,1)-\hat{q}+1)^{b(2,1)-\hat{q}+1}}{(b(2,1)-\hat{q})^{b(2,1)-
\hat{q}}}
\right ] ^{N_0(S^2(l))}=
\left [
\frac{4^4}{3^3}
\right ] ^{\frac{1}{2}N_2(S^2(l))}.
\end{eqnarray}
Thus, by setting  $C(2)\doteq[4^4/3^3]^{1/2}$ and
$C(3)\doteq[(5/2)^{5/2}/(3/2)^{3/2}]^{4/3}$, we  eventually get
\begin{eqnarray}
&&Card[B^4_{es}(S^3(1),\ldots,S^3(k);S^2(1),\ldots,S^2(k-
1))]\simeq\nonumber\\
&&\simeq\exp\left[\left (\sum_{j=1}^kN_3(S^3(j))\right )\ln{C(3)}-
\left (\sum_{l=1}^{k-1}N_2(S^2(l))\right )\ln{C(2)}\right].
\label{cardo}
\end{eqnarray}
Since
\begin{eqnarray}
\sum_{j=1}^kN_3(S^3(j))=N_3(S^3)+2\sum_{l=1}^{k-1}N_2(S^2(l)),
\end{eqnarray}
where $S^3=\partial{B^4_{es}}$ is the stacked boundary of $B^4_{es}$,
we can rewrite (\ref{cardo}) as
\begin{eqnarray}
Card[B^4_{es}(S^3(1),\ldots;S^2(1),\ldots,)]\simeq
C(3)^{N_3(S^3)}\cdot\left[ \frac{C(3)^2}{C(2)} \right]^{
\sum_{l=1}^{k-1}N_2(S^2(l))},
\label{Bentropy}
\end{eqnarray}
(by exploiting (\ref{ennesse}) this expression can be also rewritten
in terms of $N_4(B^4_{es})$). Since $C(3)/C(2)>1$, we have that
triangulations of $B^4_{es}$ with large
$\sum_{l=1}^{k-1}N_2(S^2(l))$ are dominant in the infinite volume
limit.
This implies that the simplicial volume of the $k-1$ edges connecting
the $k$ vertices is as large as possible. Note that (\ref{Bentropy})
does not depend on the particular
$S^3(j)$ or $S^2(l)$ but only on the fixed quantities $N_3(S^3)$ and
$\sum_{l=1}^{k-1}N_2(S^2(l))$ determining the ratio between $N_4(S^4)$
and the volume of the singular part
$B^4_{es}$ of ${\Bbb S}^4$, (see (\ref{sing3}) and (\ref{ennesse})).

Thus, among all possible triangulations with $k$ distinct singular
vertices
connected by $k-1$ distinct edges, those entropically favored, as $k$
varies, are the less constrained ones,
namely triangulations with just one singular edge connecting two
singular vertices:
the triangulations of $B^4_{es}$ with $k=2$. For such triangulations
the $S^3$ links of the singular vertices and the $S^2$ link of the
connecting edge are as large as kinematically possible. Note that for
the triangulated $B^4=C(\partial{B^4})$ considered in section
\ref{singvertex} we have
\begin{eqnarray}
Card[C(\partial{B^4})]\simeq
C(3)^{N_3(S^3)},
\end{eqnarray}
and in the large volume limit
\begin{eqnarray}
&&Card[B^4_{es}(S^3(1),\ldots;S^2(1),\ldots,)]\simeq\nonumber\\
&&\simeq C(3)^{N_3(S^3)}\cdot\left[ \frac{C(3)^2}{C(2)} \right]^{
\sum_{l=1}^{k-1}N_2(S^2(l))} >Card[C(\partial{B^4})]\simeq
C(3)^{N_3(S^3)}.
\label{edgesingo}
\end{eqnarray}
Since, as $N_4(S^4)$ increases, the triangulations ${\Bbb S}^4_{es}$
enter more and more
in entropic competition with the single singular vertex triangulations
${\Bbb
S}^4_{sv}$, (\ref{edgesingo})
directly implies the following basic result

\begin{lemma}
For a given ratio
\begin{eqnarray}
\frac{Vol(S^4_{es})}{Vol(sing)}=
\frac{N_4(S^4_{es})}{N_4(B^4_{es})}=\frac{22+6h}{9},
\end{eqnarray}
with $h=0,1,2,\ldots,$,
the
singular triangulations of ${\Bbb S}^4$ which are closer to the
kinematical boundary $b(4,2)=4$, and which entropically dominate in
the large volume limit $N_4(S^4)\to\infty$,
are realized by triangulations ${\Bbb S}^4_{es}$ with one sub-singular
edge connecting two singular vertices, and are characterized by the
average incidence
\begin{eqnarray}
b_h(4,2)=10\cdot\frac{22+6h}{54+15h}.
\end{eqnarray}
\label{edgelemma}
\end{lemma}

The last part of this lemma, concerning the $h$-parameterization of
the singular triangulations, is an immediate consequence of the
expressions (\ref{crumpledb2}) and (\ref{crumpledb3}) for the average
incidence of ${\Bbb S}^4_{es}$ and of the results of section
\ref{singvertex}.
Results which
characterize the sets of value of $\alpha$ and $\beta$ giving the
closest approach of
$b(4,2)=10\cdot\frac{12+2\alpha}{30+3\beta}$ to the kinematical
boundary
$b(4,2)=4$ as the ratio
$\frac{Vol(S^4_{es})}{Vol(sing)}$ varies.
\vskip 0.5 cm
The geometrical analysis just discussed and the
Lemma \ref{edgelemma} appear in good qualitative agreement with the
picture which emerges from
recent Monte Carlo simulations \cite{singedge} concerning
the study of singular structures in $4$D simplicial gravity. According
to such a numerical analysis there are, {\it at finite volume}, two
pseudo-critical couplings (and hence corresponding pseudo-critical
incidences $b(4,2)$)  separately associated with the creation of
singular edges and singular vertices. This behavior seem to correspond
to the different entropic relevance of the single singular vertex
triangulations ${\Bbb S}^4_{sv}$ and of the singular edge
triangulations ${\Bbb S}^4_{es}$ discussed above. In the simulations
the two pseudo-critical couplings lock into a single critical point in
the large volume limit. This merging appears to be related to the full
entropic competition between ${\Bbb S}^4_{sv}$ and ${\Bbb S}^4_{es}$
which dominates our geometrical picture in the infinite volume limit.
Explicitly, the average incidence
$b(4,2)|_{S^4_{es}}$, (see \ref{crumpledb3}), is slightly larger (at
finite volume) than
$b(4,2)|_{S^4_{sv}}$. Thus, if we apply formula (\ref{solution})
relating the average
incidence $b(4,2)$ to a value of the coupling $k_2$, we find that the
set of $k_2(S^4_{es})$'s corresponding to $b(4,2)|_{S^4_{es}}$, (as
$h$ varies), is slightly smaller than the corresponding set of
$k_2(S^4_{vs})$'s associated with $b(4,2)|_{S^4_{vs}}$.
Anticipating the analysis of section \ref{secquattro}, this remark
implies that there are
indeed two pseudo-critical points respectively associated with edge-
singular ${\Bbb S}^4_{es}$
and vertex-singular ${\Bbb S}^4_{vs}$ triangulations, say
$k_2^{crit}(S^4_{es};N_4)$ and
$k_2^{crit}(S^4_{vs};N_4)$, with
\begin{eqnarray}
k_2^{crit}(S^4_{es};N_4)\leq  k_2^{crit}(S^4_{vs};N_4),
\end{eqnarray}
and coalescing in just one critical point as $N_4$ gets larger and
larger. Obviously, what one actually sees at a given finite volume
mostly depends  on the rate $N_2(S^2)/N_3(S^3)$, (see
\ref{crumpledb3}), which controls how fast the two average incidences
$b(4,2)|_{S^4_{es}}$ and $b(4,2)|_{S^4_{vs}}$ approach each other. On
this rate we are not yet able to say anything substantial. As
recalled,
(see (\ref{edgexample})), computer
simulations indicates that at relatively large volumes, (tipycally
$N_4=32000$), the term $N_2(S^2)/N_3(S^3)$ is already so small that
$b(4,2)|_{S^4_{es}}\simeq{b(4,2)}|_{S^4_{vs}}$ up to a few percent,
and edge-singular
triangulations are to all effects as close to the kinematical boundary
$b(4,2)=4$ as the ${\Bbb S}^4_{vs}$ are. Thus they do entropically
dominate.

\subsection{The characterization of the critical incidence}

Since in the infinite volume limit both singular configurations ${\Bbb
S}^4_{sv}$ and ${\Bbb S}^4_{es}$ are characterized by the same average
incidence (\ref{crumpledb}), we can use indifferently both for
characterizing the critical incidence $b_0(4)$ signaling the closest
approach of generic singular triangulations to the kinematical
boundary $b(4,2)=4$.
The single singular vertex configurations ${\Bbb S}^4_{sv}$ are
somehow easier to handle
than ${\Bbb S}^4_{es}$, thus for definiteness we describe the
characterization of the critical incidence (and the corresponding
critical gravitational coupling) by referring
explicitly to
${\Bbb S}^4_{sv}\simeq{B^4}\cup_{S^3}C(\partial{B^4})$. In any case,
one should keep in mind that the extension of the analysis to ${\Bbb
S}^4_{es}$ can be carried out without difficulty along the same lines.
\vskip 0.5 cm
How can we characterize the critical incidence $b_0(4)$? A glance at
table \ref{tavola1} clearly shows that, as
$\frac{Vol(S^4)}{Vol_{sing}(\sigma^0)}$ increases,
the values of $b(4,2)|_h$ are very close to each other. This remark
implies that triangulations with
$b(4,2)|_{h=0}=110/27$, even if entropically dominating in ${\Bbb
S}^4_{sv}\simeq{B^4}\cup_{S^3}C(\partial{B^4})$, cannot be taken as
the mark of the real
critical incidence. As a matter of fact. for values of $h$ close to
the leading
configuration at $h=0$, there can be statistical competition between
such singular triangulations, at least as $N_4\to\infty$. The critical
incidence $b_0$ is actually obtained by averaging the distinct
$b(4,2)|_h$'s over the set of corresponding singular triangulations.
\vskip 0.5 cm
To characterize such average we exploit the fact that the singular
triangulations we are considering have their singular part constructed
as a cone over a stacked $3$-sphere ${\Bbb S}^3$. If we join, through
the identification of a marked $\sigma^3\in{\Bbb S}^3$, two stacked
$3$-spheres, ${\Bbb S}^3(1)$ and ${\Bbb S}^3(2)$ we get another
stacked $3$-sphere
${\Bbb S}^3(3)={\Bbb S}^3(1)\#{\Bbb S}^3(2)$, and all (voluminous)
stacked spheres can be obtained in this way. Thus, if we construct the
cone over this connected sum of stacked $3$-spheres we can sweep all
possible voluminous ({\it i.e.}, large $N_4$) singular triangulations
of the type we are considering.
Explicitly, let us denote the singular triangulations of
${\Bbb S}^4$, obtained from the stacked $3$-spheres ${\Bbb S}^3(1)$
and ${\Bbb S}^3(2)$,
by ${\Bbb S}^4(1)\doteq B^4(1)\cup_{S^3(1)}C({\Bbb S}^3(1))$
and ${\Bbb S}^4(2)\doteq B^4(2)\cup_{S^3(2)}C({\Bbb S}^3(2))$,
respectively. If
${\Bbb S}^3(3)={\Bbb S}^3(1)\#_f{\Bbb S}^3(2)$, where $f$ is an
homeomorphism between two marked
$\sigma^3(1)\in{\Bbb S}^3(1)$ and $\sigma^3(2)\in{\Bbb S}^3(2)$, then
\begin{eqnarray}
{\Bbb S}^4(3)= {\Bbb S}^4(1)\#_{f*}{\Bbb S}^4(2)=
(B^4(1)\cup_f B^4(2))\cup_{S^3(3)}C({\Bbb S}^3(3)),
\label{compo}
\end{eqnarray}
where $f*$ is the extension of $f$ to the cone over the marked
$\sigma^3$, and every singular triangulations of ${\Bbb S}^4$ over a
stacked $3$-sphere can be obtained in this way.
\vskip 0.5 cm
The analytical counterpart of (\ref{compo}) follows directly from the
last of relations (\ref{dellapalla}) characterizing the $f$-vector of
the ball $B^4$ as the parameters $\alpha$ and $\beta$, (thus $h$),
vary. From it we get
\begin{eqnarray}
N_4[B^4(1)\cup_f{B^4(2)}] &=&
\left[\frac{13+6h}{9}\right]{N}_1(S^3(1))
+\left[\frac{13+6h}{9}\right]{N}_1(S^3(2))\nonumber\\
&=&N_4[B^4(1)]+N_4[B^4(2)],
\end{eqnarray}
where we have discarded costant terms which are $o(1)$ in the large
$N_4$ limit.
To exploit this information let
\begin{eqnarray}
{\cal T}_h[Vol(B^4)=N]\doteq{Card}\left\{{\Bbb
S}^4\colon\frac{Vol_{norm}(S^4)}{Vol_{sing}(\sigma^0)}=\frac{6h+22}{9}
\
;\; Vol(B^4)=N  \right\},
\end{eqnarray}
be the  cardinality of the set of distinct  singular triangulations of
the ball
$B^4$, constructed over a stacked ${\Bbb S}^3$, with given
ratio $\frac{Vol(S^4)}{Vol_{sing}(\sigma^0)}$
and $N_4(B^4)=N$. According to the behavior of this set of
triangulations
under the connected sum we have
\begin{eqnarray}
{\cal T}_h\left[Vol(B^4)=N(1)+N(2)\right]={\cal
T}_h\left[Vol(B^4)=N(1)\right]\cdot
{\cal T}_h\left[Vol(B^4)=N(2)\right].
\end{eqnarray}
It is easily verified that this relation implies that the leading
asymptotics of ${\cal T}_h(Vol(B^4))$
is provided by
\begin{eqnarray}
{\cal T}_h(Vol(B^4))=c(B^4;h)^{N_4(B^4)},
\label{calpalla}
\end{eqnarray}
where $\ln{c(B^4;h)}$ is the specific entropy for the generic
$\sigma^4\in{\Bbb S}^4_{vs}$.

 Since there is a unique cone
$C({\Bbb S}^3)$ over the stacked sphere boundary
$\partial{B^4}\simeq{\Bbb S}^3$,
(\ref{calpalla}) provides also the leading exponential asymptotics to
the number of distinct triangulations of ${\Bbb S}^4_{vs}$ with given
$N_4$ and given $h$, {\it viz.},
\begin{eqnarray}
Card\{{\Bbb S}^4_{vs}\}\propto c(B^4;h)^{N_4(B^4)}.
\end{eqnarray}
Actually, when $h>>1$ and $N_3(S^3)=O(1)$, for each triangulation of
${\Bbb S}^3$, there can be a worth of $Aut(S^3)$ inequivalents
cones, $Aut(S^3)$ denoting the automorphisms group of the given
triangulation, for simplicity we disregard here these correction
factors. Note also that the above construction applies to
the edge-singular spheres ${\Bbb S}^4_{es}$ with minor modifications.

According to (\ref{esseffe}),
\begin{eqnarray}
N_4(B^4)=N_4(S^4)-N_3(S^3)=N_4(S^4)\frac{13+6h}{22+6h},
\end{eqnarray}
thus we get that to leading order
\begin{eqnarray}
Card\{{\Bbb S}^4_{vs}\}=c(B^4;h)^{N_4(S^4)-N_3(S^3)}\doteq
s(h)^{N_4(S^4)},
\label{ballentropy}
\end{eqnarray}
where we have introduced the specific entropy, $\ln{s(h)}$, of a
$\sigma^4\in{\Bbb S^4}_{sv}$ according to
\begin{eqnarray}
\ln{s(h)}&\doteq&\lim_{N_4(S^4)\to\infty}
\frac{\ln{Card\{{\Bbb S}^4_{vs}\}}}{N_4(S^4)}\nonumber\\
&=&\frac{13+6h}{22+6h}\ln{c(B^4;h)}.
\end{eqnarray}

In order to characterize $\ln{s(h)}$, note that triangulations of the
form ${\Bbb S}^4_{vs}$ describe, for $h=0$,
the generic singular triangulations of ${\Bbb S}^4$ realizing the
closest approach to the kinematical boundary $b(4,2)=4$. Conversely,
and as already stressed, the triangulations
${\Bbb S}^4_{vs}$ reduce, as $h\to\infty$, to the generic (branched
polymer) triangulations of ${\Bbb S}^4$, (with a rooted $\sigma^4$).
These remarks imply that corresponding to $h=0$ and
$h=h_{max}$ we must have
\begin{eqnarray}
\ln{s(h=0)}&=&\ln{c(S^4;h=0)}\nonumber\\
\ln{s(h=h_{max})} &=&\ln{c(S^4;h=h_{max})},
\end{eqnarray}
where $h_{max}$ is characterized by the value of the ratio
(\ref{critvol})  evaluated for the smallest possible $N_3(S^3)=5$,
{\it i.e.}, $h_{max}=\frac{3}{10}N_4-\frac{11}{3}$, and
where $\ln{c(S^4;h)}$ is the specific entropy
associated with the microcanonical partition function (\ref{notation})
, {\it i.e.},
\begin{eqnarray}
c(S^4;h)\simeq
\left [
\frac{(b(4,2)-2)^{b(4,2)-2}}{(b(4,2)-3)^{b(4,2)-3}}
\right ] ^{10/b(4,2)},
\label{specific}
\end{eqnarray}
with $b(4,2)=10\cdot(22+6h)/(54+15h)$, (the actual specific entropy
contains a constant
factor which is of no relevance for the present considerations-see
(\ref{notation})).

Since $c(S^4;h)$ is a slowly varying function of $h$, the specific
entropy $\ln{s(h)}$ can be characterized
as the convex combination of $\ln{s(h=0)}$ and $\ln{s(h=h_{max})}$
over the interval
$0\leq{h}\leq{h}_{max}$, {\it viz.},
\begin{eqnarray}
\ln{s(h)}=\frac{h}{h_{max}}\ln{s(h=h_{max})}+\left(1-
\frac{h}{h_{max}}\right)\ln{s(h=0)}.
\end{eqnarray}
In other words, we are considering $\ln{s(h)}$ as the convex
combination of the extreme pure phases ($h=0$: crumpling, and
$h\to\infty$: branched polymer).
A straightforward computation provides
\begin{eqnarray}
s(h)=c(S^4;h=0)\cdot\left[\frac{c(S^4;h=0)}{c(S^4;h=h_{max})}\right]^{
-\frac{10}{3N_4}h}.
\label{probo}
\end{eqnarray}
Since  in the large $N_4(S^4)$ limit,
$\ln[c(S^4;h=0)/c(S^4;h=h_{max})]\simeq0.06$ we eventually get for the
leading asymptotics
\begin{eqnarray}
Card\{{\Bbb S}^4_{vs}\}=c(S^4;h=0)^{N_4}e^{-\frac{h}{5}}.
\label{probability}
\end{eqnarray}

It is worth stressing that a completely analogous result holds for
$Card\{{\Bbb S}^4_{es}\}$,
since, as $N_4\to\infty$,
the set of edge-singular triangulations, (with one edge connecting two
singular vertices), ${\Bbb S}^4_{es}|_{k=2}$, is as close to the
kinematical boundary $b(4,2)=4$ as the
triangulations ${\Bbb S}_{vs}$. The two class ${\Bbb S}^4_{es}$ and
${\Bbb S}^4_{vs}$ only
differ in the subleading asymptotics.
\vskip 0.5 cm
According to (\ref{probability}), the average value of $b(4,2)|_h$
over the set of singular triangulations considered is given, in the
large $N_4$ limit, by
\begin{eqnarray}
\la b(4,2)_{sing}\ra|_{h_{max}}
=\frac{\sum_{h=0}^{h_{max}}b(4,2)|_h\exp[-
\frac{h}{5}]}{\sum_{h=0}^{h_{max}}\exp[-\frac{h}{5}]}.
\label{singaverage}
\end{eqnarray}
By approximating the numerator with an integral, we get
\begin{eqnarray}
\la b(4,2)_{sing}\ra|_{h_{max}}
=4+\frac{4}{15}\cdot\frac{e^{\frac{18}{25}}[E_1(\frac{18}{25})-
E_1(\frac{h_{max}}{5}+\frac{18}{25})]}{5(1-e^{-\frac{h_{max}}{5}})},
\label{accamax}
\end{eqnarray}
where $E_1(x)$ is the exponential integral function.
In the large volume limit $h_{max}\to\infty$, and the above expression
reduces to
\begin{eqnarray}
\la
b(4,2)_{sing}\ra=4+\frac{4}{75}e^{\frac{18}{25}}E_1(\frac{18}{25})
\simeq
4.0394361235.
\label{critincidence}
\end{eqnarray}
As stressed, a similar analysis carried out for the class of
singular triangulations ${\Bbb S}^4_{es}$ would provide the same $\la
b(4,2)_{sing}\ra$. It follows that, as $N_4(S^4)\to\infty$,
(\ref{critincidence}) is the value of the incidence $b(4,2)$
statistically dominating
in both sets ${\Bbb S}^4_{sv}$ and ${\Bbb S}^4_{es}$. As argued in the
previous sections,
these triangulations are the ones
characterizing the smallest possible $b(4,2)$ marking the onset of the
dominance of singular geometries. Thus, we can identify
$\la b(4,2)_{sing} \ra$ with the {\it critical} incidence $b_0$ (see
section
\ref{critica}) characterizing the transition between the weak and the
strong coupling phase of the theory, {\it i.e.},
\begin{eqnarray}
b_0(4)\doteq \la b(4,2)_{sing}\ra\simeq 4.0394361235.
\end{eqnarray}
\vskip 0.5 cm
Together with the critical incidence $\la b(4,2) \ra$ it is worthwhile
to compute the infinite volume average, over the set of singular
triangulations ${\Bbb S}^4_{sv}$ or ${\Bbb S}^4_{es}$, of the local
volume of the singular part of the triangulation, $Vol(sing)$. Note
that for the class of triangulations ${\Bbb S}^4_{sv}$,
$Vol(sing)=Vol(\sigma^0)$, whereas for the triangulations of ${\Bbb
S}^4_{es}$ dominating in the infinite volume limit, we have
\begin{eqnarray}
Vol(sing)\simeq 2Vol(\sigma^0),
\label{edgev}
\end{eqnarray}
since according to the remarks of section \ref{sedge} and lemma
\ref{edgelemma},
in such a limit,
triangulations with just two singular vertices (connected by a sub-
singular edge) dominate.

For both class of triangulations $Vol(S^4)/Vol(sing)=(22+6h)/9$, and
the
required average is provided by
\begin{eqnarray}
\la \frac{Vol(sing)}{Vol(S^4)}\ra|_{h_{max}}
=\frac{\sum_{h=0}^{h_{max}}\exp[-
\frac{h}{5}]\frac{9}{6h+22}}{\sum_{h=0}^{h_{max}}\exp[-\frac{h}{5}]}.
\label{volaverage}
\end{eqnarray}
(Strictly speaking, this ensemble average explicitly refers to the
single singular vertex triangulations ${\Bbb S}^4_{sv}$, however, as
stressed before, this ensemble average differs from the ${\Bbb
S}^4_{es}$ ensemble average by corrections which vanish as
$N_4(S^4)\to\infty$).

By approximating as usual the summatories with an integral  we get
\begin{eqnarray}
\la \frac{Vol(sing)}{Vol(S^4)}\ra|_{h_{max}} =\frac{3e^{11/15}}{10(1-
e^{-h_{max}/5})}\cdot
[E_1(\frac{11}{15})-E_1(\frac{h_{max}}{5}+\frac{11}{15})].
\end{eqnarray}

According to (\ref{edgev}), we get for the  average local volume of
the (two) most singular vertices, the  explicit expression
\begin{eqnarray}
\la Vol(\sigma^0)\ra|_{h_{max}} =
\frac{3e^{11/15}}{20(1-e^{-h_{max}/5})}\cdot
[E_1(\frac{11}{15})-E_1(\frac{h_{max}}{5}+\frac{11}{15})]\cdot{N_4},
\label{vollo}
\end{eqnarray}
which, in the infinite volume limit, reduces to
\begin{eqnarray}
\la Vol(\sigma^0)\ra =\frac{3e^{11/15}}{20} E_1(\frac{11}{15})\cdot
N_4.
\label{avertex}
\end{eqnarray}

Note that the value of the critical average incidence $\la b(4,2)
\ra\simeq4.03943\ldots$ shows that the leading configurations
contributing to the singular geometry of ${\Bbb S}^4_{es}$ are,
loosely speaking, those
for which $h\leq 6$, (see table \ref{tavola2}). Thus, a rough
indicator of what is the average
singular volume for $b(4,2)$ sufficiently smaller than $\la b(4,2)
\ra\simeq4.03943\ldots$, ({\it viz.}, when in the polymeric phase),
can be obtained by considering the average
\begin{eqnarray}
\la \frac{Vol(sing)}{Vol(S^4)}\ra|_{poly}
=\frac{\sum_{h\geq6}^{h_{max}}\exp[-
\frac{h}{5}]\frac{9}{6h+22}}{\sum_{h\geq6}^{h_{max}}\exp[-
\frac{h}{5}]}.
\end{eqnarray}
Explicitly we get
\begin{eqnarray}
\la Vol(\sigma^0)\ra_{poly} =\frac{3e^{29/15}}{20}
E_1(\frac{29}{15})\cdot N_4,
\label{branchaverage}
\end{eqnarray}
which can be interpreted as the contribution to $\la Vol(\sigma^0)\ra$
coming from the non-singular geometries in ${\Bbb S}^4_{es}$.

\section{The critical coupling $k_2^{crit}$}
\label{secquattro}
The kinematical picture which emerges from the above analysis is
immediately connected to the thermodynamical behavior of $4D$-
dynamical triangulations by recalling the results of section
\ref{canone}
according to which, as $k_2$ varies the distribution of triangulated
manifolds is strongly peaked around triangulations with an average
incidence given by $3[A(k_2)/(A(k_2)-1)]$, (see
(\ref{solution})). Thus by solving for $k_2$ the equation
\begin{equation}
\la b(4,2)_{sing}\ra=3(\frac{A(k_{2})}{A(k_{2})-1}),
\label{cappa}
\end{equation}
 we get an estimate of the value of $k_2$ corresponding to which
singular
triangulations start dominating the canonical partition function
(\ref{can}) in the {\it infinite volume limit}. Recall that singular
triangulations are those
characterizing the sub-exponential sub-leading asymptotics (see
Th.5.2.1, pp.106-118 of \cite{Carfora})
\begin{eqnarray}
\lefteqn{W[N_{2},b(4,2)]\simeq
\label{asintotica4}}\\
&&e^{(\alpha_4b(4,2))N_2}\cdot{\left [
\frac{(b-\hat{q}+1)^{b-\hat{q}+1}}{(b-\hat{q})^{b-\hat{q}}}
\right ] }^{N_{2}}
e^{[-m(b(4,2))N^{1/n_H}_4]}
{N_{2}}^{-\frac{11}{2}},
\nonumber
\end{eqnarray}
with $m(b(4,2)>0$, (see (\ref{asintotica}) for the general expression;
the above expression can be obtained from (\ref{asintotica}) by
setting $n=4$, $\alpha_2=0$, $\tau(b)=0$, and $D=0$ since we are
considering ${\Bbb S}^4$, we have also dropped a few inessential
constant terms). Thus we can identify  the $k_2$ solution of equation
(\ref{cappa}) with
the {\it critical value}, $k_2^{crit}$, of
the inverse gravitational coupling marking the transition between the
strong and weak coupling in $4D$-simplicial quantum gravity.
\vskip 0.5 cm
 Introducing in (\ref{cappa}) the values
$\la b(4,2)_{sing}\ra\simeq4.0394361235$ obtained above for the
kinematical bound
controlling the occurrence of generic singular triangulations, we get
for the critical coupling the explicit value
\begin{eqnarray}
k_2^{crit}\simeq 1.3093.
\label{cappacrit}
\end{eqnarray}

\subsection{A model for pseudo-criticality at finite $N_4(S^4)$}

It is very interesting to compare
the value for $k_2^{crit}$, already in very good agreement with what
is
found by means of Monte Carlo simulations, with the other
$k_2^{h}$'s obtained by solving equation (\ref{cappa}) with the left
member $\la b_{sing}(4,2)\ra$ replaced by the values
$b_h(4,2)$ provided by (\ref{criticalbs}). In this way we get
table \ref{tavola2}.
\begin{table}[t]
\begin{center}
\begin{tabular}{|c||c||c||c|}
h & b(2,4) & $\frac{Vol(S^4)}{Vol_{sing}(\sigma^0)}|_h$ & $k_2^h$ \\
\hline
0 & $\frac{110}{27}\simeq4.07407$ & 2.444 & $\simeq1.24465$ \\
1 & $\frac{280}{69}\simeq4.0579$ & 3.111 & $\simeq1.2744$ \\
2 & $\frac{340}{84}\simeq4.04761$ & 3.777 & $\simeq1.2938$ \\
3 & $\frac{400}{99}\simeq4.0404$ & 4.444 & $\simeq1.30746$ \\
4 & $\frac{460}{114}\simeq4.03508$ & 5.111 & $\simeq1.31762$ \\
5 & $\frac{520}{129}\simeq4.03100$ & 5.777 & $\simeq1.32545$ \\
\end{tabular}
\end{center}
\caption[tavola2]{Some of the values of $k_2^{h}$ obtained by solving
equation
(\ref{cappa}) for
$b_h(4,2)$ as $h$ varies. Such values appear strikingly near to the
values of the pseudo-critical points found in Monte Carlo simulations
as the size of the triangulations considered is increased.}
\label{tavola2}
\end{table}

According to the remarks in the previous paragraph, $k_2^{h}$,
$h=1,2,\ldots$, can be interpreted as the values of the inverse
gravitational coupling corresponding to which
the sub-leading singular configurations comes into play. In other
words, corresponding to such values of $k_2$ there are {\it distinct
peaks} in the distribution of singular triangulations of ${\Bbb
S}^4_{es}$.
The {\it leading peak} is at $k_2=k_2^{crit}\simeq1.24465$,
this corresponds to the dominance of singular triangulations for which
$\frac{Vol(S^4)}{Vol_{sing}(\sigma^0)}|_{h}=22/9$; the {\it first sub-
leading peak} occurs at
$k_2=k_2^{h=1}\simeq1.2744$, corresponding to the sub-dominance of
singular triangulations for which
$\frac{Vol(S^4)}{Vol_{sing}(\sigma^0)}|_{h}=28/9$; the {\it second
sub-
leading peak} occurs at
$k_2=k_2^{h=2}\simeq1.2938$ and is associated with the sub-dominance
of
singular triangulations for which
$\frac{Vol(S^4)}{Vol_{sing}(\sigma^0)}|_{h}=34/9$, and so on. In the
large $N_4$ limit there is enough {\it phase space} for having all
such peaks contributing to the partition function of the theory, and
the presence of the sub-dominating peaks lowers the critical incidence
from its  {\it bare} value $b(4,2)|_{h=0}$ to $\la b(4,2)_{sing}\ra$,
and
shifts the critical $k_2^{crit}$ from the {\it bare} value $1.24465$
to its effective value $k_2^{crit}\simeq1.3093$.
Using a field-theoretic image, one may say
that in the large volume limit the
fluctuations associated with the various sub-dominating peaks in the
distribution of singular triangulations dress the bare critical
incidence.

Conversely, at a finite value of $N_4$ one expects that the resulting
average $\la b(4,2)_{sing}\ra (N_4)$, computed from
(\ref{singaverage}) with
$h\leq\bar{h}(N_4)\leq{h}_{max}$, for some
$\bar{h}(N_4)$, is larger than the limiting value $\la
b(4,2)_{sing}\ra$.
Corresponding to this $\la b(4,2)_{sing}\ra (N_4)$ one gets an $N_4$-
dependent pseudo-critical point $k_2^{crit}(N_4)$ smaller than the
actual $k_2^{crit}$. Roughly speaking, at finite volume, there is no
phase space available for having all sub-dominating peaks competing
with each other according to their relative entropic relevance.
Moreover, at finite volume we should distinguish which kind of
singular geometry we are dealing with. According to lemma \ref{closer}
and (\ref{crumpledb3}), the average incidence is larger
for the edge-singular triangulations ${\Bbb S}^4_{es}$ than for the
single singular vertex triangulations ${\Bbb S}^4$. Thus,
corresponding to ${\Bbb S}^4$ or ${\Bbb S}^4_{es}$ we should
get a slightly different sequence of pseudo-critical points,
(according to (\ref{cappa}),
$k_2^{crit}(N_4)|(S^4_{es})\leq k_2^{crit}(N_4)|(S^4)$). A difference
which fades away as the volume increases.
\vskip 0.5 cm

In order to make contact with numerical simulations is worthwhile to
develop an analytical model taking care of these {\it finite size}
effects.
Again for simplicity, let us limit our analysis to the vertex singular
triangulations
${\Bbb S}^4_{vs}$, with the understanding that what we say can be
easily extended to the
edge-singular triangulations ${\Bbb S}^4_{es}$ with minor
modifications.
The starting point of our analysis is the entropic formula
(\ref{probability}) expressing, as $h$ varies, the
entropy of the triangulations ${\Bbb S}^4_{vs}$ as convex combinations
of its extreme two {\it pure phases} associated with crumpling ($h=0$)
and polymerization ($h=h_{max}\to\infty$).
Rather than use directly (\ref{probability}) we should refer to the
conditional entropy
\begin{eqnarray}
\frac{Card{\Bbb S}^4_{vs}}{Card{\Bbb S}^4}
\end{eqnarray}
which provides the contribution of the triangulations ${\Bbb
S}^4_{vs}$ to the set of all possible triangulations of ${\Bbb S}^4$,
at fixed volume.

From (\ref{probability}) and (\ref{asintotica}) we get, to leading
order in the large
$N_4(S^4)$ limit,
\begin{eqnarray}
\frac{Card{\Bbb S}^4_{vs}}{Card{\Bbb S}^4}\simeq \Omega^{N_4(S^4)}e^{-
\frac{h}{5}},
\label{condprob}
\end{eqnarray}
where $\Omega$ is the $h$-dependent constant
\begin{eqnarray}
\Omega\doteq\frac{c(S^4;h=0)}{c(S^4;h)}\simeq33.97082\cdot
\left [
\frac{(b(4,2)-2)^{b(4,2)-2}}{(b(4,2)-3)^{b(4,2)-3}}
\right ] ^{-10/b(4,2)},
\label{omecostant}
\end{eqnarray}
with $b(4,2)=10\cdot(22+6h)/(54+15h)$.

The expression (\ref{condprob}) for the conditional entropy, holds at
finite, sufficiently large $N_4(S^4)$, and, since
$\frac{Card{\Bbb S}^4_{vs}}{Card{\Bbb S}^4}\leq1$,
it implies that, at {\it finite volume}, triangulations ${\Bbb S}^4$
with
$h>>N_4(S^4)\ln\Omega$ are entropically suppressed. This remark
implies that
the configurations ${\Bbb S}^4_{vs}$,  which actually
contribute in
characterizing the critical incidence, have an entropic cut at some
value of $h$, say $\bar{h}(N_4)=O(N_4(S^4)\ln\Omega)$.
The specific entropy $\ln{c}(S^4;h)$  of $\{{\Bbb S}^4\}$ changes very
slowly with
$h$, thus at finite $N_4(S^4)\doteq{N}$, we may tentatively write

\begin{eqnarray}
\left(\frac{Card{\Bbb S}^4_{vs}}{Card{\Bbb
S}^4}\right)_N=\Omega_0^{N_4}e^{-
\frac{h}{5}},
\label{fsize}
\end{eqnarray}
for $0\leq{h}\leq\bar{h}(N)$
whereas
\begin{eqnarray}
\left(\frac{Card{\Bbb S}^4_{vs}}{Card{\Bbb S}^4}\right)_N
=\Omega_{h=h_{max}}^{N_4},
\label{polysize}
\end{eqnarray}
for $\bar{h}(N)<{h}\leq{h}_{max}$, and where $\Omega_0=1+\epsilon$,
$\epsilon>0$, is a suitable constant not differing much from $1$,
(according to (\ref{omecostant}),
$\Omega|(h=1)\simeq 1.01234$, and $\Omega|(h=10^5)\simeq1.0615$).
In other words, we are assuming that  for $0\leq{h}\leq\bar{h}(N)$ the
system may exist as
a mixture of its two extreme pure phases, whereas for
$h>\bar{h}(N)$ it collapses into its branched polymer phase. It is
worthwhile stressing that more realistically one may consider, in
place of (\ref{fsize}), a convex combination of the extreme phase
$h=0$ and the (non extreme) phase corresponding to $h=\bar{h}(N)$. By
exploitying (\ref{probo}), this prescription can be worked out without
difficulty, however it gives rise to a rather complex scaling behavior
of the resulting entropy. Moreover, the fact that $c(S^4;h)$ is a slow
varying function of $h$, makes, as we shall see,  the simpler
(\ref{fsize}) quite accurate and much easier to handle.

A qualitative characterization of $\bar{h}(N)$ as $N$ varies can be
easily obtained by the obvious scaling properties of (\ref{fsize}). If
we consider triangulations ${\Bbb S}^4_{vs}$ with two distinct
volumes, say $N_4(S^4)=N(1)$ and $N_4(S^4)=N(2)$, then
\begin{eqnarray}
\left(\frac{Card{\Bbb S}^4_{vs}}{Card{\Bbb S}^4}\right)_{N_4=N(1)}
=\left(\frac{Card{\Bbb S}^4_{vs}}{Card{\Bbb S}^4}\right)_{N_4=N(2)},
\end{eqnarray}
provided that $\bar{h}(N)$ scales with $N$ according to
\begin{eqnarray}
\bar{h}(N(2))=\bar{h}(N(1))+5[N(2)-N(1)]\ln\Omega_0.
\label{hscaling}
\end{eqnarray}
This scaling relation implies that $\bar{h}(N)$ has a linear
dependence on $N_4(S^4)$ according to
\begin{eqnarray}
\bar{h}(N_4)=5{N_4}\ln\Omega_0+\xi,
\label{accaeffetto}
\end{eqnarray}
where $\xi$ is a suitable constant. This rather simple argument does
not yet provide
the actual value of the constants $\Omega_0$ and $\xi$, however
confrontation with numerical data at $N_4(S^4)=32000$ indicates as
reliable candidates the values
\begin{eqnarray}
5\ln\Omega_0&=&\frac{1}{16000}\nonumber\\
\xi &=& -1.
\end{eqnarray}
Note that the above condition for $\Omega_0$ implies
$\Omega_0\simeq1.0000125$, a value which is perfectly consistent with
the above characterization of $\Omega$, (see (\ref{omecostant}). It
also  indicates that the triangulations ${\Bbb S}^4_{vs}$, (actually
the entropically
dominating ${\Bbb S}^4_{es}$), do
saturate the possible set of triangulations of ${\Bbb S}^4$ in the
strong coupling phase.

The $N_4$-dependent pseudo-critical incidence $\la b(4,2)_{sing}\ra
(N_4)$ and the associated
pseudo-critical point $k_2^{crit}(N_4)$ can be easily obtained from
(\ref{accamax}) by
replacing $h_{max}$ with $\bar{h}(N)$, {\it viz.},
\begin{eqnarray}
\la b(4,2)_{sing}\ra(N_4)
=4+\frac{4}{15}\cdot\frac{e^{\frac{18}{25}}[E_1(\frac{18}{25})-
E_1(\frac{\bar{h}(N)}{5}+\frac{18}{25})]}{5(1-e^{-
\frac{\bar{h}(N)}{5}})},
\label{Naccamax}
\end{eqnarray}
and by solving for $k_2$ the equation (\ref{cappa}) with $\la
b(4,2)_{sing}\ra (N_4)$ in
place of $\la b(4,2)_{sing}\ra$.
\vskip 0.5 cm

By exploiting  these results we get an overall analytic picture of the
large volume behavior of
$4$-dimensional simplicial quantum gravity which is in a surprising
agreement with the Monte Carlo simulations of the real system
\cite{singedge}.

\section{Comparison with Numerical Work}

At this stage it is indeed useful to discuss the status of our
geometrical
results in the light of the
most recent numerical work. This comparison is particularly important
since , as recalled in the introductory remarks, the current
perspective on $4$-dimensional simplicial quantum gravity has
undergone a rather drastic change. As a matter of fact, recent Monte
Carlo simulations seem to accumulate  more and more evidence for a
first order nature of the transition separating the strong and the
weak coupling regime of the theory. Taken at face value this result
suggests that dynamical triangulations is not likely to be
a viable model of quantum gravity unless one adds additional
terms to the action.
It is perhaps fair to say that the geometrical
analysis of the previous paragraphs bears relevance to
such an issue. The characterization of the critical coupling
$k_2^{crit}$ and the existence of {\it entropically sub-dominating
peaks} in the distribution of singular triangulations strongly
indicates that this geometrical picture may be responsible for the
phenomenology we see in numerical work.
\vskip 0.5 cm
Let us start by noticing that in numerical work is difficult to
resolve the various contribution
to the distribution of singular triangulations coming from the various
peaks geometrically found by our analysis. The resolving power
depends, among other parameters, on the size of the
triangulations, and as a rough indicator, the larger the size the
bigger is the set of sub-dominating singular triangulations which come
into play. Obviously the first sub-dominant terms are the most
relevant
ones, and as suggested in the previous section,
an interesting value to look at for comparison with Monte Carlo data
is the value of the inverse gravitational coupling corresponding to
the pseudo-critical average incidence
$\la b(2,4)_{sing}\ra(N_4)$. As recalled there, by solving for $k_2$
the equation
\begin{equation}
\la b(4,2)_{sing}\ra (N_4)=3(\frac{A(k_{2})}{A(k_{2})-1}),
\label{avcappa}
\end{equation}
 we obtain the value of $k_2^{crit}(N_4)$ corresponding to which we
expect to see  a clear signature of the dominance of singular
geometries in the set of triangulated $4$-spheres of volume $N_4$.
This
is actually a pseudo-critical point, the location of which depends on
$N_4$. Numerically one finds that as the {\it volume} $N_4$ of the
triangulation increases,
the corresponding pseudo-critical point $k_2^{crit}(N_4)$ increases
too,
(see {\it e.g.} \cite{Krzywicki}). Simulations and extrapolation to
triangulations with size
$N_4=48000$ and $N_4=64000$ locate the corresponding $k_2^{crit}(N_4)$
at $1.267$ and $1.273$, respectively.
\vskip 0.5 cm
According to (\ref{accaeffetto}) the actual dependence
of the number of dominating peaks, $\bar{h}(N_4)$, as a function of
the
volume $N_4$ of the triangulation, is linear according to
\begin{equation}
\bar{h}(N_4)=\frac{N_4}{16000}-1,
\end{equation}
for $N_4(S^4)\geq32000$, where the actual value,
($5\ln\Omega_0=1/16000$ and $\xi=-1$), of the constants comes from
comparison with the numerical data provided at $N_4(S^4)=32000$ by
\cite{Krzywicki}. With this expression of $\bar{h}(N_4)$
we obtain, from (\ref{avcappa})  and (\ref{Naccamax}),  the table
\ref{finitesize}.
\begin{table}[t]
\begin{center}
\begin{tabular}{|c||c||c||c|}
$N_4$ & h & Analytical $k_2^{crit}(N_4)$ & Monte Carlo
 $k_2^{crit}(N_4)$ \\ \hline
32000 & 1 & 1.25795 & 1.258 \\
48000 & 2 & 1.26752 & 1.267 \\
64000 & 3 & 1.27466 & 1.273 \\
\end{tabular}
\end{center}
\caption[finitesize]{The value of the analytical pseudo-critical
points
$k_2^{crit}(N_4)$ versus their Monte Carlo counterparts. These values
are computed  under the hypothesis that
the linear dependence of $\bar{h}(N_4)$ from
$N_4(S^4)$ is given by $h(N_4)=N_4/16000-1$.}
\label{finitesize}
\end{table}
The agreement between the analytical pseudo-critical points and the
Monte appears surprisingly good, and suggests that the identification
of our $k_2^{crit}(N_4)$ with the pseudo critical $k_2^{crit}(N_4)$
found in Monte Carlo simulations is not a mere coincidence. An
important implication of this identification, if correct, is that the
growth with $N_4$ of $k_2^{crit}(N_4)$ is due to the increasing
contribution of the sub-dominating singular triangulations. This
result provides a nice explanation to the fact that Monte Carlo data
seem to indicate that the major part of the finite size effects come
from the crumpled phase \cite{Jurke}.

By extrapolating the actual measurements, the Monte Carlo simulations
locate the critical point around $k_2^*\simeq1.327$ or around
$k_2^*\simeq1.293$, (depending if the data fit used is modeled after a
second order or a first order transition,
respectively)\cite{Krzywicki}. Again, our analytical result
$k_2^{crit}\simeq1.3093$ appears in quite a good agreement with the
numerical data, (curiously enough our $k_2^{crit}$ is, with a good
approximation, the average of the above two numerical data), and
moreover its analytical characterization provides a natural entropic
explanation to the structure and
location of the associated finite-size pseudo-critical points.

\vskip 0.5 cm
Another distinct feature of recent numerical works concerns the
bimodality
in the distribution of singular vertices seen during Monte Carlo
simulations exactly around $k_2^{crit}(N_4=32000)\simeq 1.258$,
\cite{Krzywicki}.  In this connection, particularly interesting are
the papers \cite{singedge} and \cite{Bialas}, where long run histories
(at $N_4(S^4)=32000$) provides a reliable measurement of the average
maximum vertex order near the critical point. In these simulations the
system  wanders between two states characterized by two quite distinct
values of the average maximum vertex order. In one case, this maximum
is close to $3000$, while for the other the figure is close to $1000$.
A correlation analysis shows that this metastability corresponds to
tunneling back and forth from a branched polymer state (average vertex
order
$\simeq1000$) containing no singular vertex and a crumpled state
(average vertex order $\simeq3000$) with one or two singular vertices.

According to our analysis, this behavior is the one exactly coded into
the
entropy formulae (\ref{fsize}) and (\ref{polysize}) which exactly
describe a finite
size tunnelling between a crumpled state (described by (\ref{fsize}))
and a branched
polymer state (described by (\ref{polysize}). A good indication of
the average vertex order, as we approach
the transition point for increasing $k_2$, is provided by
(\ref{avertex})
At $N_4(S^4)=32000$ this analytic formula yields
\begin{eqnarray}
\la Vol(\sigma^0)\ra_{N_4=32000} =\frac{3e^{11/15}}{20}
E_1(\frac{11}{15})\cdot (N_4=32000)\simeq3400.
\label{numvertex}
\end{eqnarray}

Conversely, if we approach the transition point by lowering $k_2$,
then a reliable indication is provided by (\ref{branchaverage}).
Explicitly we get
\begin{eqnarray}
\la Vol(\sigma^0)\ra_{poly} =\frac{3e^{29/15}}{20}
E_1(\frac{29}{15})\cdot (N_4=32000)\simeq1770.
\label{numtex}
\end{eqnarray}

Such results appear quite in reasonable agreement with the values of
$\la
Vol(\sigma^0)\ra_{N_4=32000}$ obtained during the simulations and
mentioned before.
Such data suggests
that the bimodality seen in the numerical simulation has its origin in
the presence of sub-dominating singular triangulations. In particular,
due to finite size effects
the set of subdominating singular triangulations ${\Bbb S}^4_{es}$ for
$h=0,1,\ldots,6$
seems to provide a metastable cluster of configurations that
entropically dominate the crumpled state.
\vskip 0.5 cm
Taken at face value, this set of results seem to
indicate, at least to the indulgent reader, a variety of viewpoints on
the actual status of a theoretical interpretation of the numerical
simulations: \par
{\it (I)} The bimodality as well as the implied first order
interpretation of the transition between weak and strong coupling is a
finite size effect related to: {\it (i)} The saturation of
the triangulations of $\{{\Bbb S}^4\}$ with ${\Bbb S}^4_{es}$ in the
strong coupling phase; {\it (ii)} The slow dependence of the  specific
entropy, $\ln{c}(S^4;h)$, of $\{{\Bbb S}^4\}$ from the parameter $h$
controlling the volume of the singular part of the triangulation. This
slow
variation may be responsible of the fact that the tunnelling does not
disappear as the volume of the triangulations increases. \par
Obviously, this latter remark can be easily turned inside out to
favour a less optimistic point of view:\par
{\it (II)} The slow $h$-variation in $\ln{c}(S^4;h)$ may well be such
as to mantain the bimodality
for larger and larger volumes: we have a genuine first order
transition.
\vskip 0.5 cm
It is rather clear that our analysis, being based on a sort of mean
field approximation,
cannot distinguish clearly between such two scenarios: we need sharper
entropic estimates.
Even if shamefully low in providing answers to the headlines that
numerical simulations score,
we wish to conclude with a final example pointing to a constructive
way of using
our analytical entropy estimates. This final point concerns the $k_2$
dependence of the two normalized cumulants of the distribution of the
number of vertices of
the triangulation, $c_1(N_4;k_2)$ and $c_2(N_4;k_2)$
whose analytic expression is explicit provided by (\ref{Rumulant1})
and (\ref{Rumulant2}). Strictly speaking, these expressions are
accurate only near the actual critical average incidence $b_0$,
however we can use them quite safely in a rather larger range of
variation
of $k_2$, (due to the slow variation of $b(2,4)$ as a function of
$k_2$). Accurate Monte Carlo measurements of such cumulants have been
reported in \cite{Krzywicki}, and by referring to these data for
$N_4=32000$, the comparison between MC-data and our
analytic results for  $c_1(N_4;k_2)$ and $c_2(N_4;k_2)$
are shown in table \ref{compare}.
\begin{table}[t]
\begin{center}
\begin{tabular}{|c||c||c||c||c|}
$k_2$ & $c_1(N_4;k_2)$ & $c_1(MontCarl)$ & $c_2(N_4;k_2)$ &
$c_2(MontCarl)$ \\ \hline
1.240 & 0.1935053 & 0.18970(12) & 0.109062 & 0.141(7) \\
1.246 & 0.1945674 & 0.19150(11) & 0.1194586 & 0.144(8) \\
1.252 & 0.1956271 & 0.19399(32) & 0.1465348 & 0.254(35) \\
1.258 & 0.1966846 & 0.19712(20) & 0.3996907 & 0.316(8) \\
1.264 & 0.1977398 & 0.20052(21) & 0.1844987 & 0.118(20) \\
1.270 & 0.1987927 & 0.20085(27) & 0.1274851 & 0.118(20)\\
\end{tabular}
\end{center}
\caption[compare]{A comparison between the analytical values and the
available
Monte Carlo data for the first two cumulants of the distribution of
the number of vertices of the triangulation.}
\label{compare}
\end{table}
The agreement between the analytical cumulant $c_1(k_2;N_4)$ and its
Monte Carlo counterpart is particularly good; (note that for a better
comparison with the numerical data we have actually used in
(\ref{Rumulant2}) an average between $b(4,2)|_{h=0}$ and
$b(4,2)|_{h=1}$ so as to shift from $k_2^{crit}\simeq1.3093$ to a
pseudo-critical $k_2^{crit}(N_4)\simeq1.258$).  Slightly less
impressive is the agreement between the second cumulants, but this is
to be expected since near the pseudo-critical point $k_2^{crit}(N_4)$,
the second cumulant $c_2(N_4;k_2)$ fluctuates quite wildly.
We wish to stress that such an agreement rests both on the rigorous
asymptotics (\ref{cumulant1}), (\ref{cumulant2}) and on the  scaling
{\it hypotheses}
\begin{equation}
m(k_2)=\frac{1}{\nu}|\frac{1}{b(k_2)}-\frac{1}{b_0}|^{\nu},
\label{assumption2}
\end{equation}
and
\begin{equation}
\lim_{\matrix{{\scriptstyle N_4\to\infty}\cr
{\scriptstyle k_2\to k_2^{crit}}}}
|\frac{1}{b(k_2)}-\frac{1}{b_0}|^{\nu-1}\cdot{N_4}^{\frac{1}{n_H}-
1}= \mbox{const.},
\label{scaling2}
\end{equation}
The best agreement, used in table \ref{compare}, is obtained by
choosing $\nu \approx 0.94$.
Eq.\ (\ref{assumption2}),
is nothing but a natural consequence of the vanishing of the parameter
$m(b)$ for $b(2,4)\to{b}_0$; whereas the second condition
(\ref{scaling2}) rests on a less firm ground and must be considered as
a working hypothesis to be better substantiated.
\vskip 0.5 cm
Some of the results discussed above show that the numerical evidence
pointing toward a first order nature of the transition can be
explained in a natural geometrical framework. The bimodality,
which has been underlined as a strong indication that the transition
is of a
first order, is well explained by the presence of entropically sub-
dominating peaks in the distribution of singular triangulations.
Similarly to what has been argued by Catterall et al.
\cite{Catterall},  the system tunnels among such distinct sub-dominant
configurations with some of these configurations being meta-stable
for
$N_4$ finite, (especially those with  $h\simeq0,1,\ldots$ which
dominate the crumpled phase,
and those for which $h>>1$ characterizing the branched polymer phase).
Of course the analytical
arguments provided by us are all based on a kind of mean-field
approximation, since we consider only a restricted class of
triangulations.
Mean-field analysis is in general not very reliable when it comes to
predicting the {\it order} of a phase transition. However, in this
case
we have seen that combined with an additional scaling assumption, we
get
reasonable agreement with Monte Carlo data for both
$k_2^{c}(N_4)$, $c_1(N_4)$ and $c_2(N_4)$. This might indicate
a validity beyond that usually provided by a mean-field approximation.

A good test of the reliability of the geometric truncation used
in the present work is to apply it to the more complicated
system of 4d simplicial quantum gravity coupled to Abelian gauge
fields.
In that system one seemingly observe a new interesting phase structure
\cite{new}, different from the branched polymer -- crumpled phase
originally reported in \cite{original}.

\section*{References}

\begin{description}

\bibitem[1] {Carfora}
J. Ambj\o rn, M. Carfora, A.Marzuoli,
The Geometry of Dynamical Triangulations. Lecture Notes in Phys.
{\bf m50} (Springer 1997)

\bibitem[2] {Houches}
J. Ambj\o rn, Quantization of Geometry. Lect. given at
Les Houches Nato A.S.I.: Fluctuating Geometries in
Statistical Mechanics and field Theory. Session LXII, 1994;
J. Ambj\o rn, B. Durhuus, T.J\'onsson, Quantization of Geometry
(Cambridge Monograph in Math. Phys. 1997).

\bibitem[3] {Regge}
T. Regge, Nuovo
Cim.~{\bf 19} (1961) 558.

\bibitem[4] {Frohlich}
J. Fr\"{o}hlich:
Regge calculus and discretized gravitational functional
integrals. Preprint IHES (1981), reprinted in: Non-perturbative
quantum field theory --- mathematical aspects and applications.
Selected Papers of J.Fr\"{o}hlich (World Sci. Singapore 1992)

\bibitem[5] {Catterall}
S. Catterall, G. Thorleifsson, J. Kogut, R. Renken, Nuc. Phys. B
468 (1996) 263.

\bibitem[6] {Krzywicki}
P. Bialas, Z.Burda, A. Krzywicki ,B. Petersson,
Nuc.Phys. B 472 (1996) 293. See also V.B. de Bakker:
Further evidence that the transition of
$4$D
dynamical  triangulation is 1st order. hep-lat/9603024

\bibitem[7] {singedge}
S. Catterall, R. Renken, J. Kogut, Phys.Lett.B 416 (1998) 274

\bibitem[8] {Gabrielli}
D. Gabrielli: Polymeric phase of simplicial quantum gravity, to
appear in
Phys.LettB., 1998

\bibitem[9] {Gromov}
M. Gromov, Structures m\'etriques pour les vari\'et\'es
Riemanniennes. (Conception Edition Diffusion Information
Communication Nathan, Paris 1981)

\bibitem[10]{Thurston}
W. Thurston: Shapes of polyhedra and triangulations of the sphere,
math.GT/9801088, 1998

\bibitem[11] {Dav4}
C. Itzykson,J-M. Drouffe, Statistical field theory: $p.2$.
(Cambridge University Press, Cambridge  1989)

\bibitem[12] {Walkup}
D. Walkup, Acta Math. {\bf 125} (1970) 75.

\bibitem[13] {Bialas}
S. Bilke, Z. Burda, B. Petersson, Topology in 4D simplicial quantum
gravity, in Nucl.Phys. B (Proc. Suppl.) 53 (1997) 743.

\bibitem[14] {Kuhnel}
W. K\"{u}hnel, in:
Advances in differential geometry and topology. Eds. I.S.I.- F.
Tricerri, (World Scientific, Singapore, 1990)

\bibitem[15] {Varsted}
M. Gross and D. Varsted, Nucl.Phys. {\bf B378} (1992) 367.

\bibitem[16] {Pachner}
U. Pachner, Europ.J.Combinatorics {\bf 12} (1991) 129.

\bibitem[17]{Stanley}
R. Stanley, Advances in Math. {\bf 35} (1980) 236; Also:J.
Amer. Math. Soc. {\bf 5} (1992) 805; and
Discrete Geometry and convexity, pp.212-223, Ann.
NY Acad. Sci., New York 1985

\bibitem[18] {Jurke}
J. Ambj\o rn and J. Jurkiewicz, Nucl.Phys. B 451 (1995) 643.

\bibitem[19]{new}S. Bilke, Z. Burda, A. Krzywicki, B. Petersson,
J. Tabaczek and G. Thorleifsson,
Phys.Lett. B418 (1998) 266.

\bibitem[20]{original}J. Ambj\o rn and J. Jurkiewicz,
Phys.Lett. B278 (1992) 42.

\end{description}

\end{document}